\documentclass[11pt]{article}

\usepackage{neuroai-epfl}%
\usepackage{graphicx}%
\usepackage{multirow}%
\usepackage{amsmath,amssymb,amsfonts}%
\usepackage{amsthm}%
\usepackage{mathrsfs}%
\usepackage[title]{appendix}%
\usepackage{xcolor}%
\usepackage{textcomp}%
\usepackage{manyfoot}%
\usepackage{booktabs}%
\usepackage{algorithm}%
\usepackage{algorithmicx}%
\usepackage{algpseudocode}%
\usepackage{xspace}%
\usepackage{listings}%
\usepackage{caption}%
\usepackage{natbib}

\let\citetn\citet 

\definecolor{ggreen}{HTML}{00A64F}
\definecolor{light-gray}{gray}{0.9}

\newcommand*{\modelname}[1]{{\textsc{#1}}}

\newcommand*{\ourmodel}{\modelname{Topo-Omni}\xspace}

\begin{document}

\begin{titlecard}

\papertitle{Discovering Functionally Selective Brain Regions with a Deep Topographic Multimodal Model}

\paperauthors{%
    Badr AlKhamissi\textsuperscript{*},\
    Johannes Mehrer\textsuperscript{*},\
    Lara Marinov,\
    Ahmed Abdelaal,\
    Abdulkadir Gokce,\    
    Martin Schrimpf%
}

\paperaffiliations{%
    NeuroAI Lab, EPFL\hspace{0.2cm}
    \textsuperscript{*}Equal Contribution%
}

\paperabstract{%
    Nearby neurons in cortex share similar response profiles, producing systematic spatial organization across sensory and cognitive systems. Recent topographic models reproduce aspects of this structure but remain unimodal and spatially constrain each layer separately, yielding fragmented maps that capture neither the contiguity of cortical processing streams nor their integration across modalities. 
    We introduce \ourmodel, a topographic multimodal model in which visual, auditory, and language/cognitive processing share a single contiguous in-silico sheet. Built by fine-tuning a pretrained foundation model with a spatial smoothness objective, this architecture develops clusters across modalities that are consistent with human neuroimaging, from sensory to cognitive systems.
    Driving or suppressing a cluster selectively biases or impairs perception, paralleling human intervention studies. 
    Finally, we use our model to screen for novel clusters in-silico and discover new \emph{natural landscape} and \emph{animal} networks which we validate in human data.
    A single spatial principle thus organizes representations across modalities and processing stages, yielding testable hypotheses about cortical organization.
}

\papermeta
    {\href{mailto:badr.alkhamissi@epfl.ch}{badr.alkhamissi@epfl.ch}\,\textbf{,}\
     \href{mailto:johannes.mehrer@epfl.ch}{johannes.mehrer@epfl.ch}\,\textbf{,}\
     \href{mailto:martin.schrimpf@epfl.ch}{martin.schrimpf@epfl.ch}}
    {\metalink{https://github.com/epflneuroailab/topo-omni}}
    {\metalink{https://huggingface.co/epfl-neuroai/topo-omni}}
    {\metalink{https://topo-omni.epfl.ch}}

\end{titlecard}


\section{Introduction}
\label{sec:intro}

Across species, cortex is organized as a continuous folded sheet in which nearby neurons respond to similar features. As a result, cortical computation is characterized not only by the tuning properties of individual neurons but also by the spatial arrangement of responses across the cortical surface. Decades of neuroscience research have revealed the large-scale spatio-functional organization of visual and auditory cortices: 
Visual and auditory cortices contain category-selective patches such as face-, scene-, body-, tool-, and voice-selective areas \citep{kanwisher_quest_2017, tsao_faces_2003, tsao_cortical_2006, freiwald_face_2009, pitcher_triple_2009, Pernet2015}. 
Beyond sensory processing, studies have identified higher-level networks for language processing, logical reasoning, and theory of mind \citep{Fedorenko2010, Fedorenko2013, Dufour2013}.
In short, spatial organization is a central component of cortical function across multiple levels of processing complexity.

Artificial neural networks provide powerful system-level models of cortical computation and account for substantial variance in neural and behavioral responses \citep{yamins_performance-optimized_2014, khaligh-razavi_deep_2014, Kell2018, schrimpf_brain-score_2018, schrimpf2021, mehrer_ecologically_2021, tuckute2023, tang2025many, gokce_scaling_2025, alkhamissi2025emnlp, shen2025alignment, tribe, villanueva2025predicting, schad2025vibe}. However, current brain models primarily focus on predicting functional response patterns and do not explicitly model the spatial layout of cortical activity. Recent work has begun introducing topographic models in which units are assigned positions on a two-dimensional sheet and trained with spatial regularizers, such that smooth maps and category-selective patches emerge \citep{lee_topographic_2020, keller_modeling_2021, lu_end--end_2023, margalit_unifying_2024, deb_toponets_2025, rathi_topolm_2025, Dehghani2025, Qian2026}. These models capture first spatial aspects of cortical organization and can predict certain neural and behavioral phenomena that depend on spatial structure \citep{schrimpf_topographic_2024, mehrer_model-guided_2026}.

Current topographic models have three central limitations: 
First, existing models are unimodal, focusing exclusively on either vision, audition, or language, and therefore cannot capture spatial organization in multimodal or higher-order association cortex \citep{lee_topographic_2020,keller_modeling_2021,lu_end--end_2023, margalit_unifying_2024, deb_toponets_2025, rathi_topolm_2025}.
Second, they embed each model layer on a separate two-dimensional map, producing multiple spatially disconnected in-silico sheets. This design prevents the representation of spatio-functional patterns across hierarchical levels of cortical processing, such as the progression from sensory to higher-level areas. 
Third, they are constructed via training from scratch which limits the capabilities of the model relative to more powerful pre-trained AI systems.

To address these limitations, we introduce \ourmodel, a topographic multimodal model in which major cortical networks—visual, auditory, and language/cognitive—form a single contiguous in-silico cortical sheet across all processing stages. 
This architecture supports the integration of information across modalities and enables spatial constraints to act across levels of processing complexity.
Our method enables the use of pretrained foundation models, imbuing the topographic model with state-of-the-art capabilities.

We first find that \ourmodel recapitulates the formation of category-selective regions from landmark studies in vision, audition, and language. Because imposing topography risks degrading other model properties, we then benchmark \ourmodel against suitable baselines on brain alignment and downstream task performance, showing that it remains competitive on both. Next, we validate the causal role of identified selectivity clusters through targeted ablations of units. Finally, we identify multimodal clustered networks in the pretrained model that -- to the best of our knowledge -- have not been characterized in human cortex, and evaluate these model predictions spatially on \ourmodel as well as human neuroimaging data from naturalistic movie viewing.

Taken together, our results demonstrate that cortical clustering across modalities and processing stages can emerge from spatial smoothness acting on a contiguous spatial sheet.

\begin{figure}[H]
\centering
\includegraphics[width=0.8\textwidth]{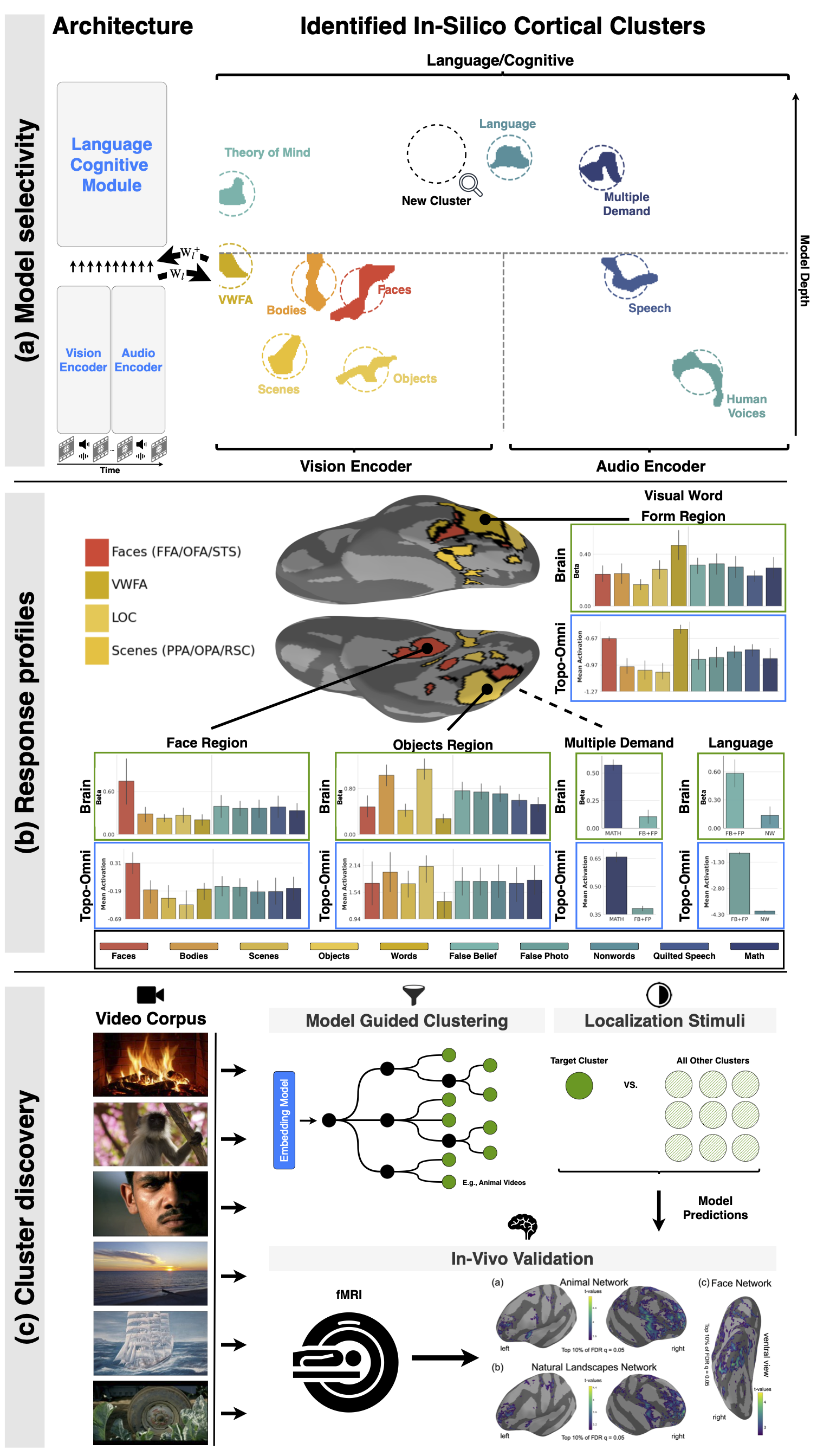}
\caption{\textbf{
(Overview) \ourmodel yields brain-like multimodal clustering.
}
\textbf{(a)} Our model builds on a multimodal architecture (left) which we project onto a single contiguous in-silico sheet (right) \ldots 
}
\label{fig:graphical-abstract}
\end{figure}

\begin{figure}[H]
  \ContinuedFloat
    \caption[]{
        (continued) 
        Selective clusters emerge in visual, auditory, and higher-level regions, from which we here visualize the largest cluster for each functional localizer  
        \textbf{(b)} Model response profiles match human fMRI responses across category-selective regions. Top: Visual clusters on the cortical surface, defined by the EMFL localizer~\cite{Marvi2025}. Bottom/right: mean response magnitudes across EMFL stimulus categories, consistent for \ourmodel{} (blue) and the human brain (green). 
        \textbf{(c)} Model-guided cluster discovery reveals novel animal- and natural landscapes-selective clusters validated in vivo. Left-to-top: Video stimuli are grouped via agglomerative hierarchical clustering on model features. Cluster-targeted localizers predict cortical responses, which are validated against the Spacetop fMRI dataset \cite{Jung2025} (bottom-right).
    }
\end{figure}

\section{Results}
\label{sec:results}

We develop \ourmodel, a multimodal model with contiguous topography across model layers. Based on \modelname{Qwen2.5-Omni} \citep{Xu2025Qwen25OmniTR}, our approach learns a bidirectional projection onto a unified cortical sheet. Governed by a self-distillation task loss and a spatial smoothness loss, this topographic arrangement spans all modalities, from vision and audio input to an integrated language/cognitive module (\href{sec:methods}{Methods}).

We evaluate \ourmodel on: 
i. the emergence of spatially localized regions corresponding to known functional systems in vision, audition, and higher cognition;
ii. the causal involvement of these regions in behavior;
and iii. whether this topographic organization is achieved without sacrificing brain alignment or multimodal task performance. 
Finally, we use \ourmodel to identify new candidate clusters and validate them using fMRI data from subjects watching movies.

\begin{figure}[H]
\centering
\includegraphics[width=1\textwidth]{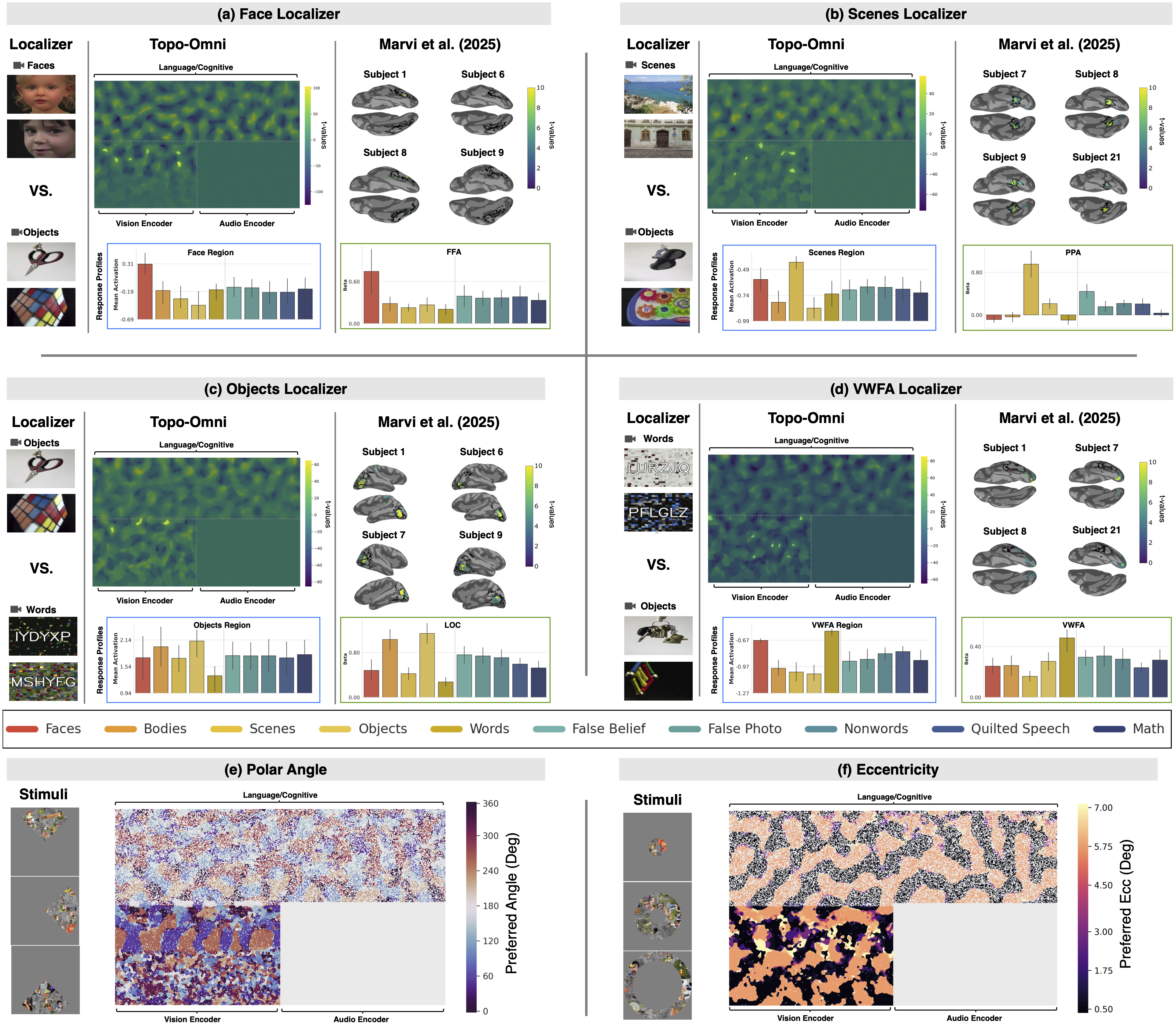}
\caption{
    \textbf{\ourmodel develops category-selective regions and retinotopic maps that parallel the functional organization of human visual cortex}. 
    Each panel shows a functional localizer contrast (videos; left), the corresponding selectivity map on the in-silico cortical sheet (centre), and analogous fMRI selectivity maps from four human subjects (right; \citetn{Marvi2025}), with response profiles of the identified region across multiple stimulus conditions for both the model and the human brain (bottom). 
    Panels test clustering for 
    \textbf{(a)} faces, 
    \textbf{(b)} scenes, 
    \textbf{(c)} objects, and
    \textbf{(d)} word form.
    In-silico and in-vivo clusters are shown across the entire respective tissue (t-values), with yellow colors corresponding to stronger preference for a contrast, and clusters falling within an anatomical localizer highlighted via black contours (Methods \ref{sec:methods:model:arch}).
    Response profiles at the bottom of each panel are shown for the average of all functionally and anatomically localized regions, with error bars as the standard deviation across stimuli or subjects.
    The bottom row test selectivity for \textbf{(e)} polar angle via rotating wedge stimuli spanning 0--360\textdegree) and \textbf{(f)} eccentricity via contracting ring stimuli spanning 0.5--7\textdegree of visual angle.
    }
    \label{fig:vision-clusters}
\end{figure}

\subsection{Emergence of visual functional organization}

We applied the multifunction human fMRI localizers from \citetn{Marvi2025} to \ourmodel in a cross-validated manner, using odd- or even-numbered runs to localize regions and held-out runs to measure response profiles (for details, see SI \S\ref{app:marvi}). This procedure recovers category-selective regions in the vision encoder that parallel the organization of the human ventral visual stream (Fig.~\ref{fig:vision-clusters}, but also see SI~\ref{sec:appendix:bodies}).

Each contrast yields spatial clustering on the simulated cortical sheet, with the in-silico response profiles positively tracking their human counterparts.
The face localizer (Faces vs.\ Objects) isolates a focal set of units that respond selectively to faces over every other stimulus category (face-selectivity $d'=0.36$; faces $>$ all other categories, paired $t(338)=27.4$, $p<0.001$, $n=339$ units). This face preference recapitulates the defining functional signature of the human fusiform face area (FFA; \citetn{kanwisher1997}). Comparing the model's response profile across the ten stimulus categories against the group-averaged FFA profile, the two are highly correlated (Pearson $r=0.88$, $p=0.012$; permutation tests; profiles averaged across units for the model and across subjects for the FFA).

The object localizer (Objects vs.\ Words) identifies clusters with elevated responses to objects and bodies (object-selectivity $d'=0.14$; objects $>$ all other categories, paired $t(300)=20.19$, $p<0.001$, $n=301$ units, top-1\%), paralleling the lateral occipital complex (LOC; Pearson $r=0.89$, $p<0.001$).

The scene localizer (Scenes vs.\ Objects) reveals a region preferring scenes (scene-selectivity $d'=0.21$; scenes $>$ all other categories, paired $t(352)=13.3$, $p<0.001$, $n=353$ units), consistent with the parahippocampal place area (PPA), though here the profile correlation reached only trend level (Pearson $r=0.63$, $p=0.077$). 

The visual word form localizer (Words vs.\ Objects) yields a region tuned to word-like stimuli (word-selectivity $d'=0.19$; words $>$ all other categories, paired $t(260)=16.9$, $p<0.001$, $n=261$ units), in line with the visual word form area (VWFA), again at trend level (Pearson $r=0.61$, $p=0.063$).

\ourmodel further shows brain-like clustering for the body localizer (Body parts vs.\ Objects; body-selectivity $d' = 0.21$, paired $t(308)=26.4$, $p<0.001$, $n=309$ units), with a cluster paralleling the extrastriate body area (EBA); here the model and human response profiles are positively but not significantly correlated (SI~\S\ref{sec:appendix:bodies}).

Across the four localized regions, model and human response profiles were positively
correlated over all ten stimulus conditions (Pearson $r=0.61$--$0.89$, mean $r=0.75$), reaching significance for FFA and LOC, and trending in the same direction for PPA and VWFA. 

Following the human fMRI analysis from \citetn{Marvi2025}, we restricted the localizer analysis in \ourmodel post-hoc to an anatomical delineation of the visual system which corresponds to the vision encoder portion of the simulated cortical sheet. Without this anatomical constraint, selective regions also emerge outside the visual system in humans, and in the language module in our model (SI~\S\ref{app:topo-clustering}). The model's audio encoder shows negligible selectivity for any of these visual localizer contrast based on video stimuli including sound regardless of whether the constraint is applied, consistent with the modality-specific selectivity of human ventral temporal cortex. 

Last, comparing \ourmodel against a non-topographic baseline trained without $\mathcal{L}_{\text{spatial}}$, the two models reproduce human ROI response profiles to a comparable degree (SI~\ref{sec:appendix:response-profiles}). 

\paragraph{\ourmodel develops retinotopic maps that parallel the organization of early human visual cortex.}
Early visual cortex is organized retinotopically: neighboring cortical units respond to neighboring positions in the visual field, producing smooth gradients of preferred polar angle and eccentricity. We tested whether \ourmodel{} captures this lower-level organizational principle using standard retinotopic mapping stimuli: rotating wedges for polar angle and contracting rings for eccentricity, and computed each unit's preferred parameter as the stimulus eliciting its maximal response (Figure~\ref{fig:vision-clusters}e-f), with units falling below a response threshold masked out (for details, see $\S\ref{subsec:receptive-field}$). Despite no supervision for retinotopy, \ourmodel{} develops continuous, smoothly varying maps for both polar angle and eccentricity in a subset of the sheet, with adjacent units sharing similar visual-field preferences. However, differences to biology remain: no clear pinwheels emerge, and there is no increase in eccentricity across processing stages. We suspect this might be due to a lack of anatomical markers in our model.  

The emergence of retinotopy alongside category-selective regions indicates that the spatial smoothness objective recovers known aspects of cortical organization across the full visual hierarchy, from low-level position coding to high-level category specialization.


\subsection{Emergence of auditory functional organization}

\begin{figure}[H]
\centering
\includegraphics[width=1\textwidth]{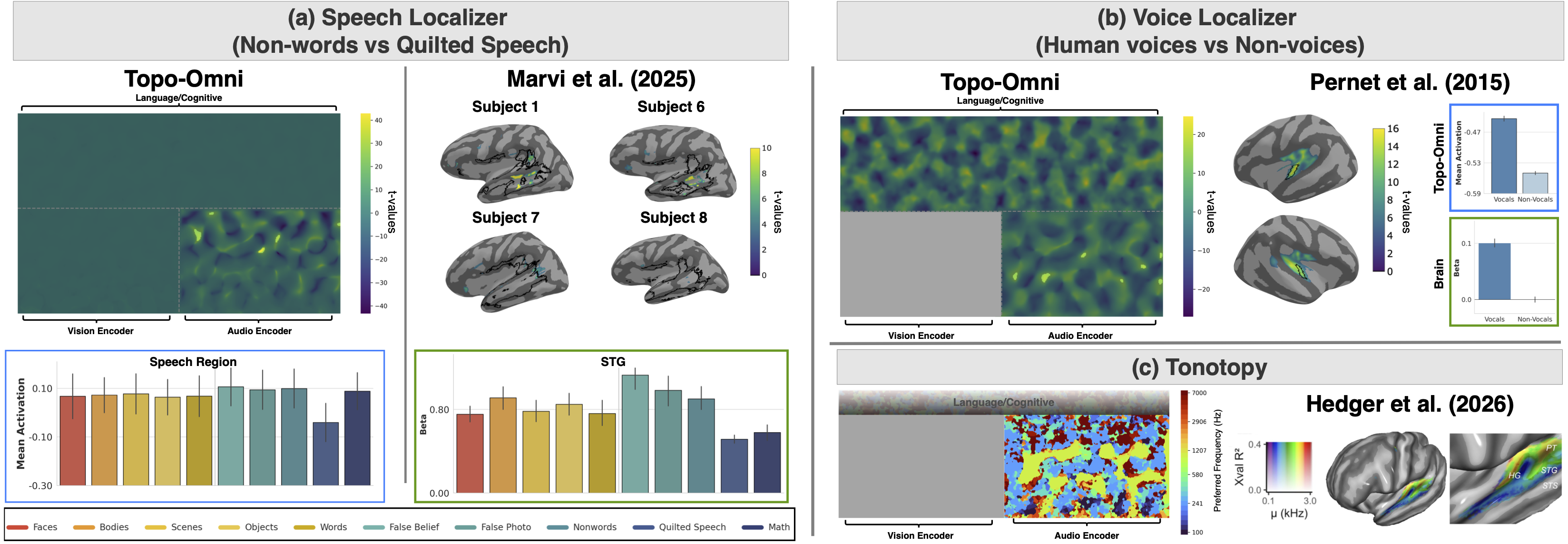}
\caption{
    \textbf{Topo-Omni develops functional organization that parallel the human auditory cortex.} 
    \textbf{(a)} Speech localizer stimuli (Non-words vs. Quilted Speech) show speech-selective regions in \ourmodel and the brain. 
    In-silico (left) and human fMRI (right) activation maps across the entire tissue, with yellow colors indicating contrast selectivity, clusters within an anatomical localizer highlighted via black contours, and non-selective regions in grey. Localizer stimuli and human data (n=6) are from \citetn{Marvi2025}.
    Response profiles (bottom) show the average across model clusters, next to human superior temporal gyrus (STG, for details, see \S\ref{app:marvi}).
    \textbf{(b)} Voice localizer stimuli (Human Voices vs. Non-voices) show human-voice-selective regions in \ourmodel and the brain (reproduced from \citetn{Pernet2015}). Maps as in (a), with response profiles based on a cross-validated analysis inspired by \citetn{Marvi2025} (SI section \ref{apx:pernet_cross_validated_response_profile}). As this localizer uses exclusively auditory input, activations are confined to the audio encoder and the language/cognitive module of \ourmodel, with no input to the vision encoder (shown in grey).
    \textbf{(c)} Tonotopy. In-silico (left) and human (right; reproduced from \citetn{Hedger2026}) maps of preferred frequency, with color indicating the frequency that elicits each unit's maximal response (SI \S\ref{app:Hedger}). The model's audio encoder develops a spatially organized preferred-frequency map, with neighboring units sharing similar best frequencies. As in (b), this analysis uses exclusively auditory input, so frequency tuning is confined to the audio encoder, with no input to the vision encoder (shown in grey).
}
\label{fig:audio-clusters}
\end{figure}

We next applied auditory localizers to \ourmodel and recovered speech- and voice-selective regions in the audio encoder that parallel the organization of human auditory cortex (Fig.~\ref{fig:audio-clusters}).

The speech localizer (Non-words vs.\ Quilted Speech, again drawn from \citetn{Marvi2025}, for details, see SI \S\ref{app:marvi}), isolates a region in the audio encoder whose response profile mirrors that of the human superior temporal gyrus (STG; Pearson $r=0.69$, $p=0.025$; permutation test). Both the model region and the human STG respond broadly across conditions containing intelligible speech and show a reliable drop for quilted speech, in which the speech signal is destroyed while low-level auditory features are preserved (quilted speech $<$ all other conditions, $d'=-0.19$, paired $t(339)=-40.5$, $p<0.001$, $n=340$ units in the considered ROI). The selectivity therefore reflects sensitivity to speech structure rather than to acoustic energy alone. 

The voice localizer (Human Voices vs.\ Non-voices), adapted from \citetn{Pernet2015} (for details, see SI \S\ref{app:pernet}), yields a distinct voice-selective region in the audio encoder that responds preferentially to human speech stimuli (words, syllables or sentence extracts from 4 languages) over non-speech sounds (laughs, sighs, cries, or coughs), paralleling the temporal voice area identified along the superior temporal sulcus in human listeners. As this localizer uses audio-only input, there is no activity in the vision encoder (shown in gray).

Beyond category selectivity, the audio encoder additionally develops a spatially organized map of preferred frequency that is similar to the tonotopic organization of human auditory cortex (Fig.~\ref{fig:audio-clusters}c, for details, see SI \S\ref{app:Hedger}).

Mirroring the findings of visual functional localizers, these regions are preferentially confined to the audio encoder portion of the cortical sheet. The vision encoder shows no selectivity for the speech localizer even though there is visual input, while the language module exhibits clustering for voices but not speech.

For comparisons of the degree of clustering in \ourmodel vs. its non-topographic counterpart, please see SI~\S\ref{app:topo-clustering}. The model's vision encoder shows no selectivity for the speech localizer contrast based on auditory stimuli with concurrently presented videos, consistent with the modality-specific selectivity of human auditory cortex.

\subsection{Emergence of higher cognitive networks}

\begin{figure}[H]
\centering
\includegraphics[width=1\textwidth]{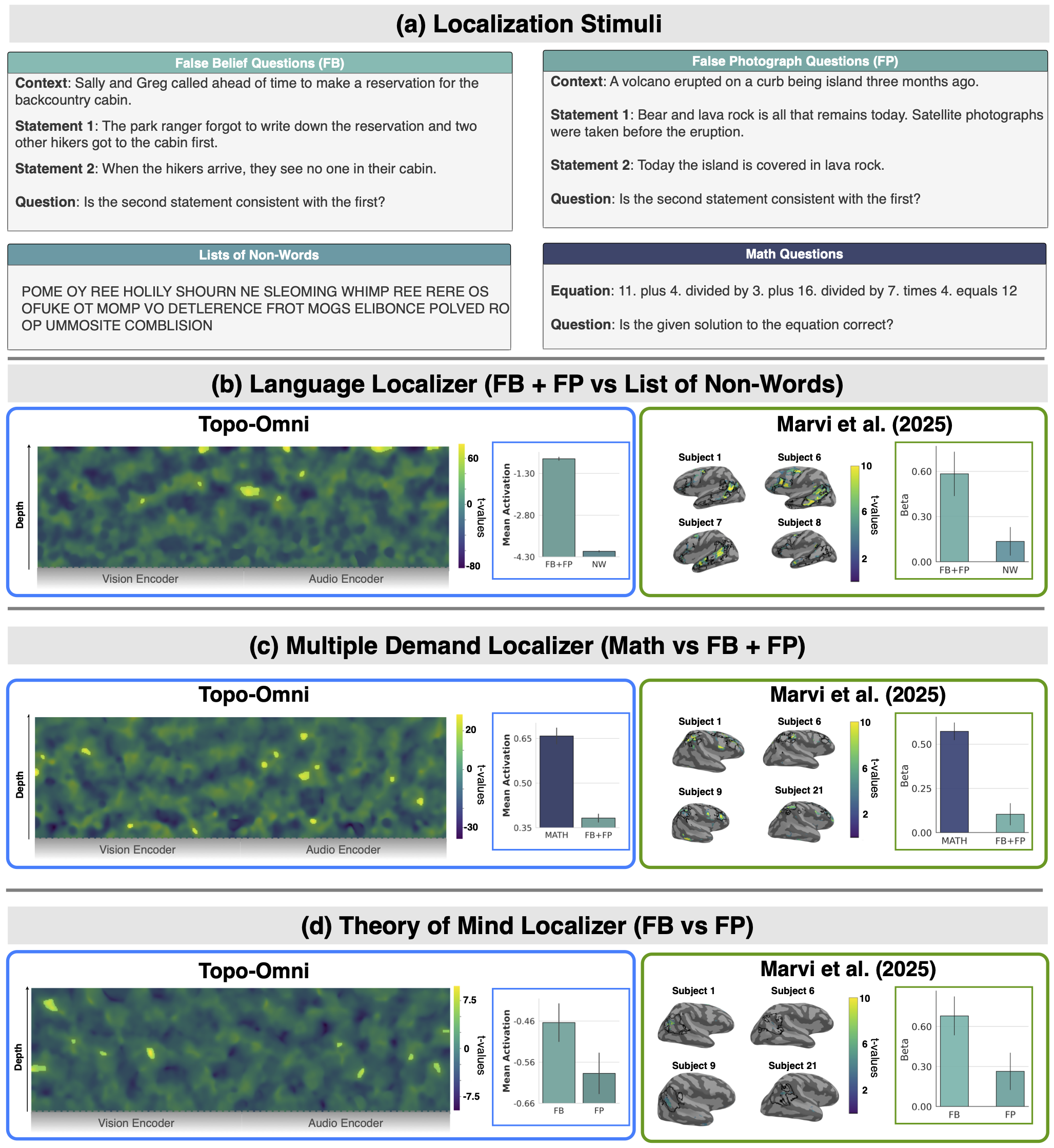}
\caption{
    \textbf{\ourmodel develops spatially distinct cognitive-task-selective networks that parallel the functional organization of human cognitive networks.} Each panel shows the activation map on the \ourmodel language module, alongside analogous fMRI activation maps from human subjects \citep{Marvi2025} in response to the same sets of stimuli. Localizer stimuli \textbf(d) are input to the model as text tokens directly. Response profile barplots are averaged across an entire network in the model or brain. Error bars are across stimuli for the model and subjects for the neural data.
    Panels test clustering for 
    \textbf{(a)} language (, Sentences vs. Non-words), 
    \textbf{(b)} multiple demand (Math Questions vs. False Belief and False Photograph Questions), 
    \textbf{(c)} theory of mind (False Belief Questions vs. False Photograph Questions).
}
\label{fig:cognition-clusters}
\end{figure}

We next asked whether \ourmodel develops spatially organized regions selective for cognitive functions. Because the higher cognitive function localizers from \citetn{Marvi2025} operate on linguistic stimuli, we passed the inputs directly as text tokens to the language/cognitive module (bypassing the vision and audio encoders thus circumventing sensory input) and examined the resulting activation maps on the simulated cortical sheet (Fig.~\ref{fig:cognition-clusters}, for details, seeSI~\S\ref{app:marvi}). Three classical cognitive localizers each isolate a selectivity network in \ourmodel that parallels their human counterparts.

The language localizer \citep{Fedorenko2010} contrasts intact English sentences with lists of non-words that preserve phonotactic structure but carry no semantic content. This contrast reveals a language-selective network in the simulated tissue that responds strongly and preferentially to meaningful sentences (language-selectivity $d'=1.39$, paired $t(621)=28.1$, $p<0.001$, $n=622$ units), paralleling the distributed fronto-temporal language network in human cortex.

The multiple demand localizer \citep{Fedorenko2013} contrasts multi-step arithmetic problems against narrative questions involving social or physical inference but minimal computational load. A multiple demand-selective network emerges in-silico that respond preferentially to the mathematically demanding condition (selectivity $d'=0.54$, paired $t(585)=34.9$, $p<0.001$, $n=586$ units), mirroring the frontoparietal multiple demand network that activates broadly during cognitively demanding tasks in humans.

Finally, the theory of mind localizer \citep{Dufour2013} contrasts False Belief questions, which require reasoning about others' mental states, with False Photograph questions, which require comparable logical inference about outdated physical representations but no mentalizing. This contrast isolates a mentalizing-selective network in the model, albeit with weaker selectivity than the language and multiple-demand regions (selectivity $d'=0.15$, paired $t(597)=25.4$, $p<0.001$, $n=598$ units), consistent with the human Theory of Mind network across the temporo-parietal junction and medial prefrontal cortex.


Together with the visual and auditory results, these findings suggest that a single contiguous topographic objective is sufficient to recover brain-like modality- and function-appropriate organization in-silico: category-selective visual regions in the vision encoder, speech- and voice-selective regions in the audio encoder, and language, multiple demand, and theory of mind networks in the language/cognitive module.
The emergence of all these clusters is driven entirely by the combination of task optimization with the simple spatial smoothness optimization (Methods~\S\ref{sec:methods:spatial-training}; Fig.~\ref{fig:methods}), with no brain data or category labels supplied during training.

\subsection{High functional brain alignment and task performance}

A central concern with imposing a spatial smoothness objective is that it may distort the model's representations, such that either its alignment with neural data or its downstream task performance is degraded. To test this, we compared three models: \ourmodel (trained with a task and spatial loss jointly), \modelname{Qwen2.5-Omni-3B SFT} (the same backbone fine-tuned with the task loss only, without the spatial term), and the original \modelname{Qwen2.5-Omni-3B} baseline.

We measured brain-model alignment on the Natural Scenes Dataset \citep{allen2021nsd} using linear predictivity (for methodological details, see $\S$\ref{sec:methods:brain_alignment}). For each model, we selected the top 10\% of units whose selectivity best matched the functional ROIs defined by the localizer stimuli of \citetn{Marvi2025}, and evaluated their ability to predict held-out fMRI responses across face-, scene-, body-, and word-selective regions in the human ventral stream. Across the twelve ROIs, the three models achieve nearly identical noise-corrected Pearson correlations (Table~\ref{tab:results}). To test whether \ourmodel diverges from either baseline, we ran two-sided paired $t$-tests across subjects within each ROI (\ourmodel vs.\ each baseline, uncorrected). The difference is not significant in 11 of the 12 ROIs (all $p > 0.05$); the sole exception is Occipital Word Form Area (OWFA), where the difference, though statistically significant, is negligible in magnitude ($\leq 0.005$ in Pearson's $r$). \ourmodel thus matches baselines across regions, indicating that the topographic constraint does not compromise the encoder's ability to predict human brain activity.

We further evaluated model task performance on OmniBench \citep{li2025omnibench}, a multimodal benchmark that requires jointly interpreting image, audio, and text inputs. \ourmodel achieves the best overall accuracy and the best performance on the Sound Event subtask, and stays within one percentage point of the SFT baseline on Music and Speech (Table~\ref{tab:results}). Because all three models are graded on the same items, we assessed each per-subtask difference between \ourmodel and the baselines with McNemar's exact test on the discordant predictions; none reaches significance (all $p > 0.05$). Together, these results demonstrate that imposing a single contiguous topographic objective on \modelname{Qwen2.5-Omni-3B}, thus producing \ourmodel, preserves both its neural predictivity and its multimodal task competence: the spatial organization recovered in the previous sections comes at no measurable cost to either alignment or downstream capability.

\begin{table*}[t]
\centering
\small
\setlength{\tabcolsep}{3.5pt}
\caption{\textbf{Brain alignment and multimodal task performance.} \textit{Top:} Brain-Score results (Pearson's $r$, noise-corrected) on the Natural Scenes Dataset \citep[NSD;][]{allen2021nsd}, reported as mean $\pm$ std across subjects. For each model, we selected the top-10\% of units whose selectivity best matched the functional ROIs defined by the localizer stimuli of \citetn{Marvi2025}, then evaluated brain-model alignment via linear predictivity ($\S$\ref{sec:methods:brain_alignment}). The $p$ columns report two-sided paired $t$-tests of Topo-Omni against each baseline across subjects (\textsuperscript{n.s.}~=~$p>0.05$; $^{*}~=~p<0.05$, $^{**}~=~p<0.01$, $^{***}~=~p<0.001$; uncorrected); Topo-Omni is not significantly different in 11 of 12 ROIs. \textit{Bottom:} Results on OmniBench \citep{li2025omnibench}, evaluating multimodal understanding across simultaneous image, audio, and text inputs, with accuracy reported overall and per audio input type. \textbf{Bold} indicates the best model per row.}
\label{tab:results}
\begin{tabular}{llccccc}
\toprule
 & & \textbf{Topo-Omni} & \textbf{SFT-Omni} & \textbf{Qwen2.5-3B} & \textbf{$p$ (SFT)} & \textbf{$p$ (Base)} \\
\midrule
\multicolumn{7}{l}{\textit{Brain-Score (NSD): Pearson's $r$, mean $\pm$ std}} \\
\midrule
Faces & OFA & $.662 \pm .03$ & $.662 \pm .03$ & $.658 \pm .03$ & $0.508^{\mathrm{n.s.}}$ & $0.056^{\mathrm{n.s.}}$ \\
 & FFA-1 & $.742 \pm .02$ & $.740 \pm .02$ & $.738 \pm .02$ & $0.314^{\mathrm{n.s.}}$ & $0.067^{\mathrm{n.s.}}$ \\
 & FFA-2 & $.732 \pm .03$ & $.730 \pm .03$ & $.730 \pm .03$ & $0.234^{\mathrm{n.s.}}$ & $0.239^{\mathrm{n.s.}}$ \\
\midrule
Scenes & OPA & $.743 \pm .03$ & $.744 \pm .03$ & $.743 \pm .03$ & $0.376^{\mathrm{n.s.}}$ & $0.910^{\mathrm{n.s.}}$ \\
 & PPA & $.821 \pm .03$ & $.820 \pm .03$ & $.821 \pm .03$ & $0.698^{\mathrm{n.s.}}$ & $0.564^{\mathrm{n.s.}}$ \\
 & RSC & $.778 \pm .03$ & $.779 \pm .03$ & $.777 \pm .03$ & $0.532^{\mathrm{n.s.}}$ & $0.510^{\mathrm{n.s.}}$ \\
\midrule
VWFA & OWFA & $.593 \pm .03$ & $.597 \pm .03$ & $.598 \pm .03$ & $0.024^{*}$ & $<0.001^{***}$ \\
 & VWFA-1 & $.689 \pm .03$ & $.689 \pm .03$ & $.689 \pm .03$ & $0.958^{\mathrm{n.s.}}$ & $0.983^{\mathrm{n.s.}}$ \\
 & VWFA-2 & $.679 \pm .03$ & $.684 \pm .03$ & $.679 \pm .03$ & $0.275^{\mathrm{n.s.}}$ & $0.797^{\mathrm{n.s.}}$ \\
\midrule
Bodies & EBA & $.740 \pm .03$ & $.739 \pm .03$ & $.739 \pm .03$ & $0.341^{\mathrm{n.s.}}$ & $0.539^{\mathrm{n.s.}}$ \\
 & FBA-1 & $.692 \pm .03$ & $.692 \pm .03$ & $.691 \pm .03$ & $0.883^{\mathrm{n.s.}}$ & $0.834^{\mathrm{n.s.}}$ \\
 & FBA-2 & $.742 \pm .03$ & $.739 \pm .03$ & $.741 \pm .03$ & $0.123^{\mathrm{n.s.}}$ & $0.774^{\mathrm{n.s.}}$ \\
\midrule
\multicolumn{7}{l}{\textit{OmniBench: accuracy (\%)}} \\
\midrule
OmniBench & Overall & 43.78 & 43.70 & 43.35 & $1.000^{\mathrm{n.s.}}$ & $0.405^{\mathrm{n.s.}}$ \\
 & Music  & 52.83 & 53.77 & 52.83 & $1.000^{\mathrm{n.s.}}$ & $1.000^{\mathrm{n.s.}}$ \\
 & Speech & 43.58 & 43.84 & 43.06 & $0.839^{\mathrm{n.s.}}$ & $0.454^{\mathrm{n.s.}}$ \\
 & Sound  & 40.75 & 39.25 & 40.38 & $0.125^{\mathrm{n.s.}}$ & $1.000^{\mathrm{n.s.}}$ \\
\bottomrule
\end{tabular}
\end{table*}

\subsection{Causal control of visual perception in \ourmodel}
\label{sec:causal}

\begin{figure}[H] 
\centering
\includegraphics[width=1\textwidth]{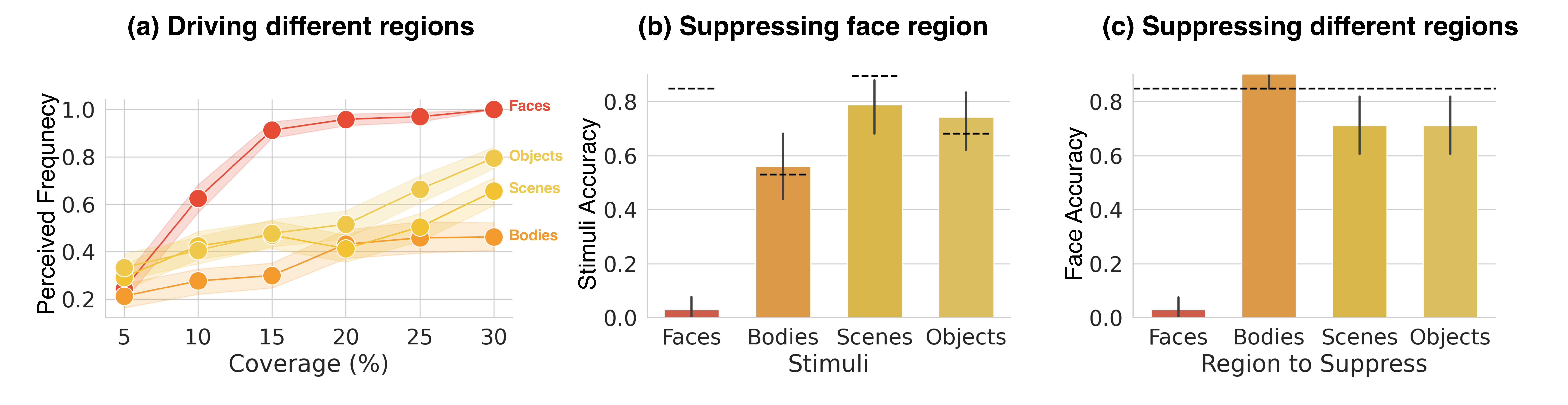}
\caption{
    \textbf{Face-selective regions in Topo-Omni enable causal control of visual perception.} 
    \textbf{(a)} Biasing perception (category frequency, $y$) by increasing activity in category-selective regions (coverage, $x$; \S\ref{sec:methods:causal}). Error bars throughout represent standard deviation across stimuli.
    \textbf{(b)} Categorization accuracy of different stimuli when suppressing the top 10\% of face-selective units in the vision encoder; classification accuracy reported separately per stimulus category, relative to baseline (dashed bars). 
    \textbf{(c)} Face identification performance when suppressing different category-selective regions (Faces, Bodies, Scenes, and Objects); dashed bar indicates baseline accuracy. 
    }
\label{fig:causal-intervention}
\end{figure}

The localizer analyses above establish that \ourmodel develops spatially clustered, category-selective regions that resemble those in human cortex, with intact behavioral outputs. 
We asked whether these category selective areas in the model are causally involved in perception, or whether they are merely correlated readouts of computations distributed across the model. We focus on the face region as the clearest test case, and report analogous interventions for the other categories.

\paragraph{Driving: face-selective units are sufficient to induce face perception.} We tested whether artificially activating a category-selective region is sufficient to bias the model's perception toward that category, regardless of the actual input stimulus (Methods~\S\ref{sec:methods:causal}). Driving an increasing fraction of units within the face region produces a sharp rise in face identification, with the model reporting face perception for nearly all stimuli when 15\% of all face-selective units are steered towards the ``face'' direction (Fig.~\ref{fig:causal-intervention}a). Body, scene, and object regions show qualitatively similar but much weaker effects.

\paragraph{Suppressing: face-selective units are necessary for face perception.} 
We suppressed the top 10\% of face-selective units in the vision encoder of \ourmodel and compared classification accuracy across four stimulus categories to the unperturbed model. Face identification collapses to near-zero, while accuracy on bodies, scenes, and objects is largely preserved (Fig.~\ref{fig:causal-intervention}b). This effect is cluster-specific: suppressing body-, scene-, or object-selective regions instead leaves face identification almost entirely intact, with only modest reductions relative to the unperturbed model (Fig.~\ref{fig:causal-intervention}c). 
The face regions in \ourmodel are thus necessary for face perception and selective in their role: face recognition depends on this topographically localized cluster and not on category-selective machinery elsewhere on the sheet.

Together, the suppression and driving experiments demonstrate that the category-selective regions in \ourmodel are not epiphenomenal: they are functionally specialized circuits that are both necessary and sufficient for category-level perception, mirroring the causal structure observed in biological visual cortex via TMS and intracranial stimulation studies \citep{pitcher_triple_2009, tsao_faces_2003, tsao_cortical_2006}. Importantly, while causal interventions can in principle be applied to any model \citep{alkhamissi2025-localization}, topography makes them anatomically targeted: the spatial smoothness objective yields clusters compact enough to be stimulated or suppressed as coherent units, mirroring the spatially localized perturbations used in human and animal neuroscience. This stands in contrast to the distributed selectivities typical of standard, non-topographic artificial neural networks, where no such localized target exists.

\subsection{Model-guided discovery of novel cortical selectivity networks}

Beyond reproducing established functional regions, we asked whether \ourmodel{} could further be used to discover category-selective clustering not yet characterized in humans. We developed an algorithm (Alg.~\ref{alg:cluster-discovery}, Sec.~\ref{subsubsec:hierarchical_clustering}) that combines agglomerative hierarchical clustering (Ward's linkage) over semantic embeddings of video segments with selectivity profiles across the simulated cortical sheet. For each resulting network, we visualized its selectivity on the model's cortical sheet and identified the video segments in a video fMRI dataset \citep[Spacetop;][]{Jung2025} that elicited the highest selectivity scores. These segments defined model-derived contrasts, which we then evaluated on human fMRI data from the same dataset (Fig.~\ref{fig:graphical-abstract}c).

\begin{figure}[H] 
\centering
\includegraphics[width=1\textwidth]{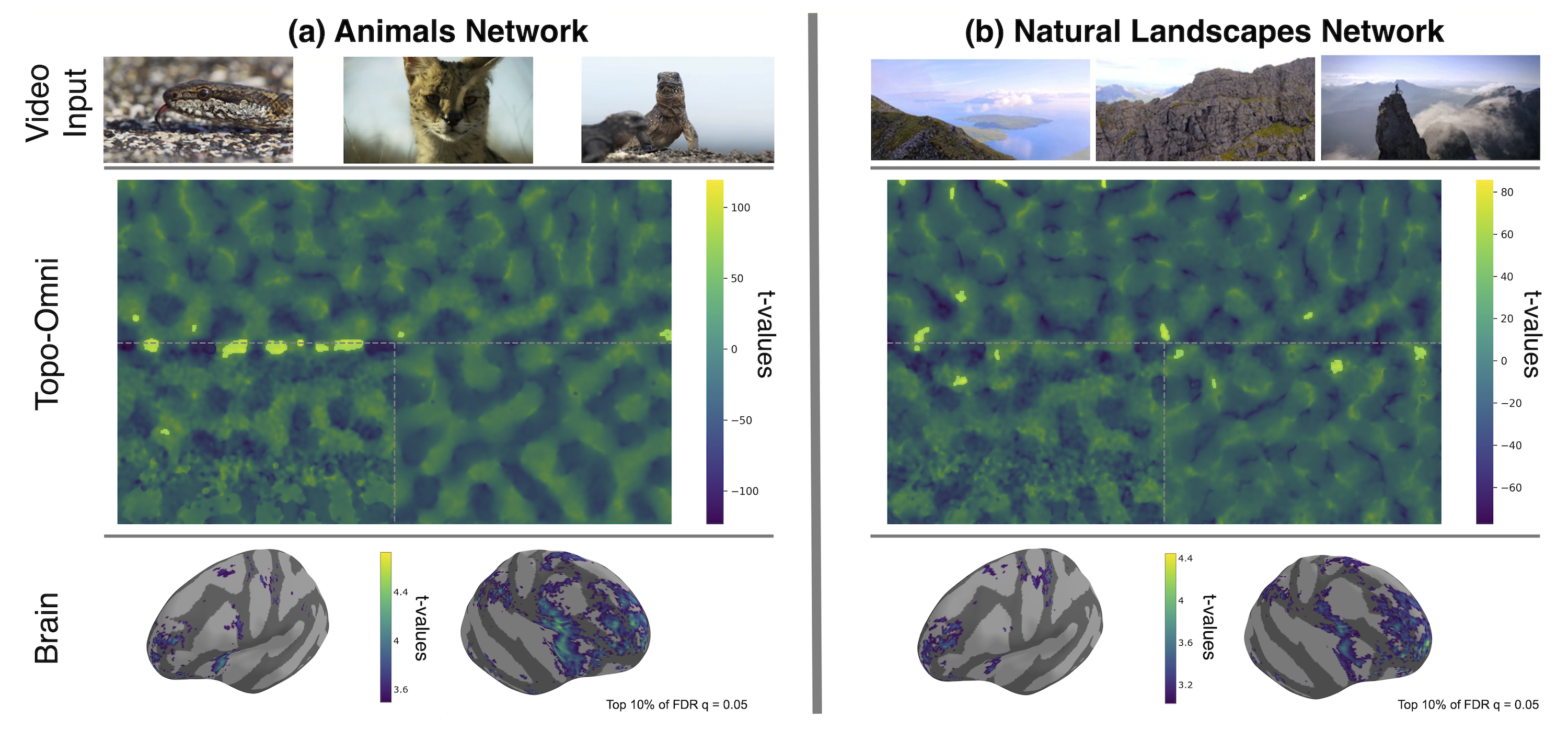}
\caption{
    \textbf{Model-guided discovery of novel cortical networks.}
    Localizer stimuli (video segments) are derived via agglomerative hierarchical clustering of \ourmodel{}'s cortical sheet, and tested on human fMRI data (Spacetop). 
    Each panel shows sample video frames (top), the \ourmodel top-1\% selectivity map (middle), and the corresponding activation map in the brain for both hemispheres (bottom; thresholded at top 10\% of FDR-significant voxels).
    Model-discovered and fMRI-validated networks selective for animals (\textbf{a}; e.g. snake, eagle, langur) and nature scenes (\textbf{b}; outdoor natural landscapes, e.g. beach, rocky terrain, alpine peaks).
}
\label{fig:cluster-discovery}
\end{figure}

This procedure recovered three reliable networks (Fig.~\ref{fig:cluster-discovery}, SI~\S\ref{app:Jung}). The first responded selectively to faces, with cortical responses concentrated in ventral visual cortex in close proximity to the canonical face-processing network (Fig.~Appendix \ref{fig:apx:cluster_exploration_netowork_brain_maps_1}). This acts as a positive control confirming that the pipeline recovers known selectivity using a model-guided data-driven approach. The remaining two networks, to our knowledge, are not described in the existing literature: one selective for \textit{animals}, including snakes, birds, and primates (Fig.~\ref{fig:cluster-discovery}a); and one for \textit{natural landscapes}, including beaches, rocky terrain, alpine landscapes (Fig.~\ref{fig:cluster-discovery}b). Human brain activity in response to these video segments validates the model predictions, with right-lateralized clustering in prefrontal cortex at the top 10\% of FDR-significant voxels ($q < 0.05$; one-tailed Welch's $t$-test). Additional frames for all three clusters, together with fMRI maps for the faces cluster, are provided in SI~\S\ref{app:Jung}. Prior work contrasting indoor and outdoor scenes found greater parahippocampal activation for indoor scenes and no region preferring outdoor (natural) scenes \cite{henderson_cortical_2007}. Our contrast instead isolates natural content against diverse non-scene categories and reveals a prefrontal network.

\section{Discussion}
\label{sec:discussion}

\ourmodel demonstrates that a single organizing principle is sufficient to produce brain-like multimodal and multi-stage functional organization. Specifically, spatial smoothness over a contiguous cortical sheet yields cortical clustering in visual, auditory, and higher-cognitive systems within one model. The clusters that emerge are not only spatially and selectively aligned with their human counterparts but are causally implicated in the model's behavior and predictive enough to guide the discovery of novel cortical organization in human cortex. Together, these results move topographic modeling beyond demonstrations of emergent maps in single sensory domains toward a unified framework in which the same spatial constraint organizes representations across modalities and processing stages.

\paragraph{Spatial smoothness as a general organizing principle.}
Prior topographic artificial neural networks have shown that imposing spatial smoothness can give rise to brain-like functional organization, first in visual cortex \citep{lee_topographic_2020, keller_modeling_2021, lu_end--end_2023, margalit_unifying_2024} and more recently in language \citep{rathi_topolm_2025, deb_toponets_2025}. Across these models, the spatial smoothness objective serves as an efficient computable proxy for wiring-length minimization, encouraging units with similar response profiles to co-localize and reducing the need for long-range connectivity. Whereas previous models impose spatial smoothness on a separate sheet per layer or modality, \ourmodel embeds visual, auditory, and higher-cognitive units in a single contiguous sheet, allowing the same wiring-cost principle to organize representations both within and across processing stages. The recovery of visual, auditory, and higher-cognitive clusters and networks under this unified objective suggests that the wiring-cost principle is not domain-specific but may capture a general organizing pressure for cortical-style spatial layout. We note that other biologically plausible drivers -- such as developmental gradients, input statistics, intrinsic connectivity priors, and learning dynamics -- likely contribute to cortical organization alongside wiring-cost considerations. Distinguishing their relative roles remains an open question that \ourmodel does not adjudicate.

\paragraph{Interpretation of the in-silico cortical sheet.}
\ourmodel captures organizational principles, not fine-grain anatomy. The in-silico cortical sheet is an abstracted substrate from the biological implementation: it does not model hemispheres, cortical folding, cytoarchitecture, or the precise relative positions of human functional regions, and it does not distinguish anatomical subregions within a broader category-selective class (e.g., face-selective regions in human cortex OFA vs. FFA). 
While a simple spatial smoothness objective applied across a multimodal computational substrate is sufficient to recover the spatio-functional organization observed in human cortex at the level of category-selective regions, stronger anatomical correspondences would require correspondingly stronger architectural priors, which we see as an exciting avenue for future work.

\paragraph{Causal interventions as a methodological capability.}
A central advantage of spatially localized functional specialization is that it makes causal interventions spatially interpretable. In \ourmodel, localizer-defined units form compact category-selective regions whose activations can be suppressed or driven to mimic the spatially targeted perturbations used in human and animal neuroscience. We illustrate this in the face-selective region: suppressing it selectively abolishes face identification while leaving other categories intact, and driving it biases the model toward face responses regardless of the actual input. We note that functional localizers in non-topographic models allow for similar ablations \citep{alkhamissi2025-localization}. 

Our results based on \ourmodel indicate that emergent clusters are causally implicated circuits. Beyond serving as a check on the model, this property enables in-silico analogues of TMS, intracranial stimulation, and lesion studies -- interventions that in humans and animals are scarce, costly, or infeasible to run at scale. \ourmodel can therefore be used to screen for causally involved regions that can be spatially interpreted, before running in-vivo experiments \citep{mehrer_model-guided_2026}. 

\paragraph{Model-guided discovery of cortical organization.}
Beyond recovering known functional areas, \ourmodel provides a framework for model-guided discovery of cortical organization. By clustering naturalistic video segments using selectivity derived from \ourmodel (or a non-topographic variant of \ourmodel) and then testing the resulting contrasts in human fMRI, we identified candidate animal- and nature-selective clusters predominantly in prefrontal cortex. 
To our knowledge, these have not previously been described as functionally selective regions in the same sense as classical face-, place-, word-, voice-, or language-selective areas.
This closed loop from in-silico clustering to predicted contrast to in-vivo validation illustrates a mode of neuroscience in which models propose hypotheses about cortical organization that are subsequently tested in humans, rather than serving only as post-hoc accounts of existing findings \citep{yamins_using_2016,richards_deep_2019,schrimpf_integrative_2020,doerig_neuroconnectionist_2023}.

\paragraph{Topography preserves alignment and task performance.}
Imposing spatial organization could in principle degrade either neural alignment or task competence, which has been documented in prior work \cite{lee_topographic_2020, margalit_unifying_2024}. Our comparisons to non-topographic Qwen2.5-Omni-3B variants demonstrate that this is not the case: \ourmodel matches or exceeds the baselines on brain predictivity across twelve NSD ROIs and on downstream task performance in terms of OmniBench accuracy. Spatial organization therefore need not be treated as a biological detail that trades off against computational performance and can be incorporated into high-performing multimodal systems at no measurable cost.

\paragraph{A platform for spatially grounded NeuroAI.}
Several research directions follow from our approach. First, the contiguous sheet enables model-guided localizer design: clusters identified in the model can propose further contrasts to be tested in independent human or animal experiments, complementing the conventional pipeline in which models account for already-discovered regions. Second, the causal intervention machinery enables in-silico screening before TMS or intracranial stimulation studies, generating predictions about which regions are necessary or sufficient for specific perceptual or cognitive outcomes \cite{mehrer_model-guided_2026}. Third, the multimodal structure invites systematic study of cross-modal organization at component boundaries, where the same spatial loss can pull together units representing semantically related content across vision, audio, and language. Fourth, the architectural template generalizes: any multimodal foundation model can in principle be fitted with a contiguous topographic sheet using the projection scheme introduced here, opening the door to topographic variants spanning additional modalities such as touch, olfactory, or motor processing. We view \ourmodel as a first instance of this broader class of models.

\paragraph{Limitations.}
Several limitations remain. 
First, as discussed above, the in-silico cortical sheet captures coarse organizational principles rather than detailed anatomy, and does not model hemispheric organization, cortical folding, white-matter connectivity, cytoarchitecture, or the precise relative positions of human functional regions. 
Second, our validation of known localizer responses uses the publicly available subset of the EMFL dataset, containing only a limited number of participants (n=6). 
Third, \ourmodel is trained on approximately $\sim$4,500 videos, which is modest at this model scale. The behavior of the spatial loss under substantially larger training corpora remains an open question. 
Fourth, our self-distillation training paradigm of using the unmodified Qwen2.5-Omni-3B baseline's outputs as targets preserves capability but couples the spatial loss to a specific functional anchor. Whether comparable organization emerges under training from scratch or under alternative task objectives is a direction for future work. 
Fifth, our novel-cluster findings rest on a single dataset (Spacetop), a single statistical pipeline, and no causal validation in humans. Establishing that the predicted prefrontal regions are causally involved in animal- or nature-related processing will require independent stimulus sets, independent subject samples, and intervention experiments such as TMS.

\subsection{Conclusion}
\label{sec:conclusion}

\ourmodel provides evidence that a spatial smoothness principle can induce brain-like spatio-functional organization across visual, auditory, and higher-cognitive domains within a single topographic multimodal model. By embedding multiple processing stages and modalities in a contiguous in-silico cortical sheet, \ourmodel extends topographic ANN modeling beyond isolated unimodal maps and converts the model into a platform for generating spatially and causally testable hypotheses about cortical organization. 
More broadly, our results suggest that ANN-based brain models can move beyond accounting for known neural responses and begin to predict previously uncharacterized functional organization in cortex.


\section{Methods}
\label{sec:methods}

\begin{figure}[H] 
\centering
\includegraphics[width=1\textwidth]{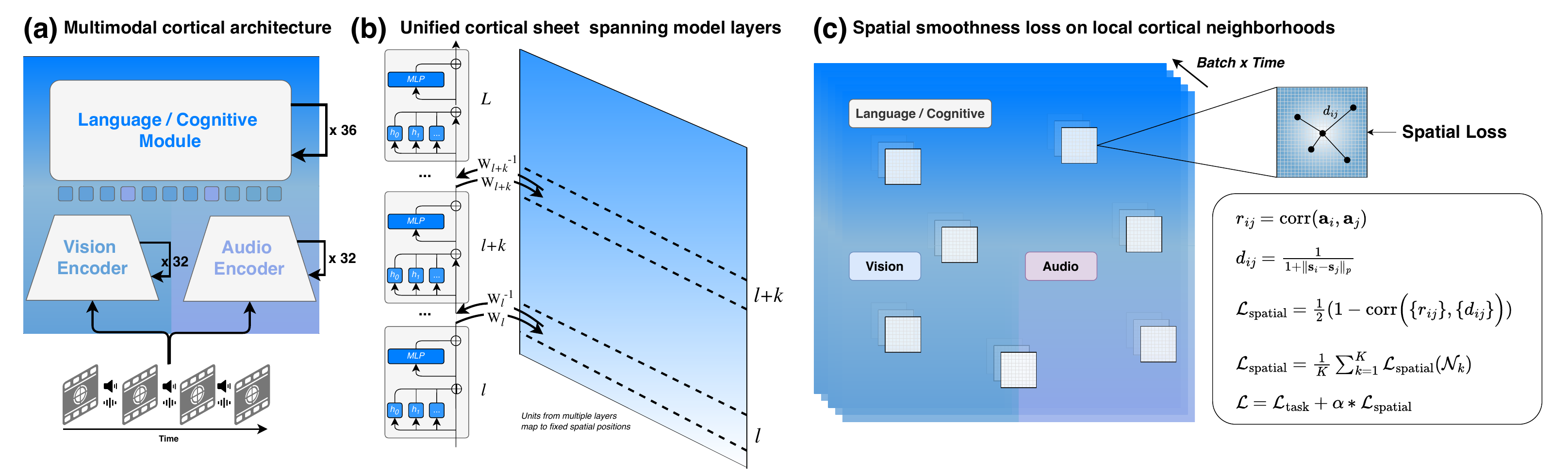}
\caption{
        \textbf{Topo-Omni: a topographic multimodal model with a unified cortical sheet.}
        \textbf{(a)} Multimodal base model architecture. The model consists of a Vision Encoder (32 layers), an Audio Encoder (32 layers), and a Language/Cognitive module (Thinker, 36 layers), which integrates visual, auditory, and text tokens to produce the final response.
        \textbf{(b)} Unified cortical sheet spanning model layers. Intermediate activations from each Transformer layer are reshaped into rectangular sheets and stacked along the depth dimension. Units from multiple layers map bidirectionally to fixed spatial positions, forming one contiguous cortical sheet per component. The vision and audio sheets are placed side by side and the language/cognitive sheet is positioned on top, yielding a single two-dimensional sheet for the entire model.
        \textbf{(c)} Spatial smoothness loss on local cortical neighborhoods. For randomly sampled neighborhoods on the unified sheet, pairwise functional similarity ($r_{ij}$, Pearson correlation of activation vectors across the batch) is aligned with pairwise spatial proximity ($d_{ij}$, inverse distance on the sheet). The total training objective combines the standard task loss with the spatial smoothness loss: $\mathcal{L} = \mathcal{L}_{\text{task}} + \alpha \, \mathcal{L}_{\text{spatial}}$.
}
\label{fig:methods}
\end{figure}

\subsection{Model}
\label{sec:methods:model}

We use the pretrained \modelname{Qwen2.5-Omni-3B}\footnote{\url{https://huggingface.co/Qwen/Qwen2.5-Omni-3B}} as our base model and fine-tune it with a spatial loss and a self-distillation task loss, as described in \S\ref{sec:methods:spatial-training}. This yields \ourmodel, a topographic multi-modal model that can predict the presence of functional clusters in the human brain from auditory, visual, or audiovisual stimuli. 

\subsubsection{Architecture}
\label{sec:methods:model:arch}

The \modelname{Qwen2.5-Omni-3B} architecture consists of three major components: the \textit{Vision Encoder} processing visual input, the \textit{Audio Encoder} processing auditory input, and the \textit{Thinker}, which we refer to as the \textit{Language/Cognitive} module. This module integrates the outputs of the vision and audio encoders (and optionally direct text-token input) to produce the final response (Fig.~\ref{fig:methods}a).

\paragraph{Vision Encoder}
The vision encoder maps images (video frames) to a sequence of continuous embeddings that can be consumed by the language model. Concretely, the input is resized and split into fixed-size patches, which are linearly projected into a patch-embedding space and augmented with positional encodings. A stack of Transformer blocks (ViT architecture~\citep{dosovitskiy2021vit}) then produces contextualized visual tokens. These visual tokens are finally projected into the shared multimodal embedding space used by the \textit{language/cognitive module}, so that vision features can be fused with audio and language features via attention.

\paragraph{Audio Encoder}
The audio encoder maps a raw waveform to a sequence of acoustic embeddings. The waveform is first transformed into a time--frequency representation (e.g., output of log-mel filterbanks), after which a learnable front-end and a stack of Transformer-based layers produce contextualized audio tokens. Similar to vision, the resulting audio tokens are projected into the shared multimodal embedding space, enabling direct cross-modal fusion in the \textit{language/cognitive module}.

\paragraph{Language/Cognitive Module}
This module is a decoder-only Transformer that integrates the modality-specific tokens (visual and/or audio) with text tokens. It performs cross-modal reasoning by attending over the concatenated token sequence and produces the final response autoregressively. In our work, we treat the \textit{language/cognitive module} as the main computational substrate for multimodal integration. 

\subsubsection{Unified Cortical Sheet}
\label{sec:methods:model:unified-sheet}

Each component described above is a Transformer with residual connections between layers (Fig.~\ref{fig:methods}). To introduce spatial structure, we insert a trainable linear projection $W_l$ after each Transformer layer that maps every token's intermediate activation onto a fixed-size two-dimensional sheet, with sheet dimensions independent of sequence length (Fig.~\ref{fig:methods}b). The sheet contains as many units as the layer's hidden dimension $d$, so $W_l$ is square and is initialized near the identity, $W_l = I_d + E$ with $E_{ij} \sim \mathcal{N}(0, 10^{-6})$ i.i.d. ($\sigma=10^{-3}$). This near-identity start preserves the pretrained representation at initialization while allowing the spatial objective to gradually reshape the mapping during training. We then project back to the residual stream using the pseudo-inverse $W_l^{+}$, ensuring that activations are minimally perturbed when returned to the model and that the forward pass is functionally preserved up to the sheet's rank. Crucially, this routing removes the original direct connections between Transformer layer  and instead forces the computational graph to pass through the cortical sheet itself, enabling the causal interventions described in \S\ref{sec:causal}. We then concatenate the per-layer sheets along one spatial axis to form a single larger two-dimensional sheet for each component, with layer index running along the concatenation axis.

We assemble the three components into a single unified sheet by placing the vision and audio encoder sheets side by side and positioning the \textit{language/cognitive} sheet on top. This arrangement is the geometric basis for the contiguous topographic objective: because all three components share one continuous sheet, the spatial loss can span the entire model rather than being applied independently per component or layer, as in existing topographic models. Finally, we average the sheets across token timesteps in windows of two seconds, thus matching the repetition time (TR) of fMRI acquisition. From the resulting unified sheet, we sample patches to compute the spatial loss as described in \S\ref{sec:methods:spatial-training}.

\subsection{Spatial Smoothness Loss on a Unified Cortical Sheet}
\label{sec:methods:spatial-training}

\ourmodel induces brain-like spatio-functional organization by optimizing a spatial smoothness objective over the unified cortical sheet defined in \S\ref{sec:methods:model:unified-sheet}, which spans the vision encoder, audio encoder, and language/cognitive module. This objective encourages nearby units on the sheet to exhibit similar response profiles, providing a differentiable proxy for minimizing neural wiring cost, following prior work on topographic deep neural networks \citep{lee_topographic_2020, margalit_unifying_2024, rathi_topolm_2025}. 

\subsubsection{Units on the unified cortical sheet}
We define the set of units
\[
\mathcal{U} = \{u_1, \dots, u_N\}
\]
as the positions on the unified cortical sheet, obtained by projecting each Transformer layer's activations through the trainable map $W_l$ described in \S\ref{sec:methods:model:unified-sheet}. Because the vision encoder, audio encoder, and language/cognitive module share a single sheet by construction, every unit $u_i$ is assigned a distinct and fixed coordinate
\[
\mathbf{s}_i \in \mathbb{R}^2
\]
in a common two-dimensional coordinate system, regardless of which component it originates from.

Given an input batch, each unit $u_i$ produces an activation vector
\[
\mathbf{a}_i \in \mathbb{R}^{B\times T},
\]
where $B$ is the number of videos in the batch and $T$ is the number of two-second chunks per video (one per fMRI TR; \S\ref{sec:methods:model:unified-sheet}). Pearson correlations between units are then computed over the $B \times T$ stimulus contexts as follows.


\subsubsection{Pairwise functional similarity and spatial proximity}
For a sampled subset of units $\mathcal{U}' \subset \mathcal{U}$, we compute pairwise functional similarity using Pearson correlation:
\[
r_{ij} = \mathrm{corr}(\mathbf{a}_i, \mathbf{a}_j),
\qquad i \neq j,\; u_i, u_j \in \mathcal{U}'.
\]
Spatial proximity is defined as a monotonically decreasing function of distance on the cortical sheet:
\[
d_{ij} = \frac{1}{1 + \|\mathbf{s}_i - \mathbf{s}_j\|_\infty},
\]
so that nearby units have higher proximity values and distant units have lower values. We use the $\ell_\infty$ norm throughout.

\subsubsection{Spatial smoothness objective}
The spatial loss encourages alignment between functional similarity and spatial proximity. For a sampled unit set $\mathcal{U}'$, we define
\[
\mathcal{L}_{\text{spatial}}(\mathcal{U}')
= \frac{1}{2}
\left(
1 - \mathrm{corr}\bigl(\{r_{ij}\}, \{d_{ij}\}\bigr)
\right),
\]
where the correlation is computed over all unordered pairs $(i,j)$ with $u_i, u_j \in \mathcal{U}'$ and the factor $\tfrac{1}{2}$ rescales the loss to $[0,1]$. Minimizing this loss encourages nearby units to develop correlated response profiles while allowing distant units to vary more freely, leading to the emergence of spatially contiguous functional clusters.

\subsubsection{Practical computation: neighborhood sampling}
Computing the spatial loss over all $O(N^2)$ unit pairs is too expensive. Following prior topographic models \citep{margalit_unifying_2024, rathi_topolm_2025}; we approximate the objective using local cortical neighborhoods. At each training step, we sample $K$ neighborhoods $\{\mathcal{N}_k\}_{k=1}^K$, where each $\mathcal{N}_k$ is the set of units within a fixed spatial radius around a randomly chosen anchor location on the unified sheet (Fig.~\ref{fig:methods}c). The spatial loss is computed independently within each neighborhood and averaged:
\[
\mathcal{L}_{\text{spatial}}
= \frac{1}{K} \sum_{k=1}^{K}
\mathcal{L}_{\text{spatial}}(\mathcal{N}_k).
\]
This local approximation enforces smoothness at the scale of cortical neighborhoods without imposing global constraints, while remaining computationally tractable. We set $K=100$ neighborhoods when training \ourmodel{}.

\paragraph{Cross-modal organization.}
Because anchor locations are sampled uniformly across the unified sheet, neighborhoods can straddle the boundaries between the vision encoder, audio encoder, and language/cognitive module. The same spatial loss therefore drives both intra-modal clustering of functionally similar units and coherent cross-modal organization at component boundaries, allowing shared functional representations to co-localize across vision, audio, and language, a property that is structurally inaccessible to prior topographic models trained on a single component in isolation.

\subsection{Task Loss and Training Data}
\label{sec:methods:task-loss}

The spatial smoothness objective alone provides no signal about what the model should compute; without a task constraint, minimizing $\mathcal{L}_{\text{spatial}}$ would freely distort the learned representations and degrade both neural alignment and downstream task performance. We therefore train \ourmodel with a joint objective
\[
\mathcal{L} = \mathcal{L}_{\text{task}} + \alpha\,\mathcal{L}_{\text{spatial}},
\]
where $\mathcal{L}_{\text{task}}$ is a supervised fine-tuning (SFT) loss that anchors the model to the capabilities of its Qwen2.5-Omni-3B initialization. We set $\alpha = 20$.

\paragraph{Training data.} We compiled a dataset of 4{,}364 videos sampled from Koala-36M \citep{wang2024koala36m}. For each video, we generated a caption by prompting the un-modified Qwen2.5-Omni-3B baseline with a question drawn from a pool of diverse captioning prompts (e.g., \textit{``What is shown in this video?''}, \textit{``What action is taking place?''}). Using the baseline model's own outputs as targets ensures that the SFT loss pulls \ourmodel toward the behavior of its pre-trained initialization rather than introducing a distributional shift from an external annotation source.

\paragraph{Task objective.} $\mathcal{L}_{\text{task}}$ is the standard cross-entropy loss computed on the assistant tokens of each (video, prompt, caption) triple. This self-distillation setup acts as an anchor: it preserves the multimodal capabilities of the baseline model while leaving the spatial loss free to reorganize representations on the cortical sheet. As shown in Table~\ref{tab:results}, this functional anchoring is effective in practice: \ourmodel matches the brain predictivity of Qwen2.5-Omni-3B on the Natural Scenes Dataset and matches or exceeds its performance on OmniBench, confirming that the topographic constraint can be imposed at no measurable cost to either neural alignment or task competence.

\subsection{Human Neural Responses}
\label{sec:methods:neural}

\subsubsection{Vision, Audio, Higher-level Cognition: \citep{Marvi2025}}
\label{sec:methods:vision}

We analyzed the publicly available subset of the Efficient Multifunction fMRI Localizer (EMFL) dataset from \citetn{Marvi2025}. This subset contains fMRI data from 6 participants who completed 5 runs of an approximately 14-minute blocked localizer experiment. Participants viewed videos from five visual categories (faces, bodies, scenes, objects, and words on scrambled backgrounds) while simultaneously listening to auditory or cognitive stimuli from five categories (false-belief stories, false-photo stories, nonwords, quilted speech, and arithmetic problems). The orthogonal combination of visual and auditory/cognitive streams allowed us to estimate responses to all 10 stimulus conditions within a single GLM and to compute the 9 functional contrasts used by \citetn{Marvi2025} to localize 13 functional regions spanning visual, speech, language, theory-of-mind, and multiple-demand systems. We used these contrasts to define functional ROIs and extract cross-validated response profiles, as described in detail in SI~\S\ref{app:marvi}.

\subsubsection{Audio: High-level Auditory Areas \citep{Pernet2015}}
\label{sec:methods:pernet}

We analyzed the publicly available temporal voice area fMRI dataset from \citetn{Pernet2015} (n=218). Participants passively listened to vocal sounds, non-vocal sounds, and silence blocks. We used the vocal $>$ non-vocal contrast to identify voice-selective temporal cortex, following \citetn{Pernet2015}. Full preprocessing, GLM, group-level analysis, and visualization details are provided in SI~\S\ref{app:pernet}.

\subsection{Measuring Model-Brain Alignment}
\label{sec:methods:brain_alignment}

\subsubsection{Natural Scenes Dataset}
\label{sec:methods:brain_alignment_nsd}

The Natural Scenes Dataset (NSD) \citep{allen2021nsd} is a large-scale 7T fMRI dataset in which eight participants viewed thousands of natural images over repeated scanning sessions while performing a continuous recognition task. The dataset provides single-trial response estimates in several spatial representations and preprocessing variants. In all NSD analyses, we use the released \texttt{b3} beta estimates, which combine voxel-wise hemodynamic response modeling, GLMdenoise \citep{kay2013glmdenoise}, and ridge regularization, and we use data in the subject-native \texttt{func1pt8mm} volumetric space.

For each subject, we restrict analyses to voxels within the released \texttt{nsdgeneral} mask and retain voxels that pass a $10\%$ noise-ceiling threshold. Noise ceilings are computed from the provided \texttt{ncsnr} files following the NSD release and prior work. Within this visually responsive mask, we analyze only a subset of functionally localized ROIs corresponding to four higher-level functional domains. Specifically, we use the following ROI groups: \textbf{Faces}: \texttt{OFA}, \texttt{FFA-1}, and \texttt{FFA-2}; \textbf{Visual word form area (VWFA)}: \texttt{OWFA}, \texttt{VWFA-1}, and \texttt{VWFA-2}; \textbf{Scenes}: \texttt{OPA}, \texttt{PPA}, and \texttt{RSC}; and \textbf{Bodies}: \texttt{EBA}, \texttt{FBA-1}, and \texttt{FBA-2}. These ROI groups define the super-categories used for model-unit localization and final score aggregation.

Each NSD subject viewed a large subject-specific set of unique images as well as a separate shared set of 1{,}000 images that was repeated across all participants. We use the subject-specific unique images as the training set and the shared image set as the held-out test set. Voxel responses are z-score standardized within session, and responses are averaged across available stimulus repetitions. Because several subjects did not complete all scan sessions, some images have fewer than three repetitions; in these cases, we average across all available repetitions.

\subsubsection{Brain Alignment, Functional Localization, and Aggregation}
\label{sec:methods:brain_alignment}

We quantify model--brain alignment by asking how well functionally localized model units predict measured neural responses under a standardized encoding-model pipeline. For each model, let
\[
\mathcal{U} = \{u_1, \dots, u_M\}
\]
denote the full set of candidate units used for alignment. In topographic models, these units are taken from the model's cortical sheet. In the baseline model, we replace the projection $W_l$ with an identity function. For a stimulus $\mathbf{x}$, the model yields an activation vector
\[
\mathbf{z}(\mathbf{x}) \in \mathbb{R}^{M},
\]
where each entry corresponds to the response of one candidate unit in $\mathcal{U}$.

In our main NSD analyses, we do not fit encoding models on all candidate units at once. Instead, we first perform an independent functional localization of model units, as described in \citetn{alkhamissi2025-localization}, and rank all units by their selectivity for each localizer $g \in \{\text{faces}, \text{vwfa}, \text{scenes}, \text{bodies}\}$. We retain the top-$10\%$ of units for ROI $g$, denoted $\mathcal{U}_{g}^{(p)} \subset \mathcal{U}$. The restricted feature representation is then
\begin{equation}
\label{eq:selected_features}
    \widetilde{\mathbf{z}}_{g}^{(p)}(\mathbf{x})
    =
    \mathrm{select}\!\left(\mathbf{z}(\mathbf{x}), \mathcal{U}_{g}^{(p)}\right).
\end{equation}
This procedure tests whether the units identified by functional localization are especially predictive of the corresponding cortical regions.

For each subject $s$ and ROI $r$, we fit a linear readout from the selected model features to the measured voxel responses,
\begin{equation}
\label{eq:linear_readout_selected}
    \widehat{\mathbf{y}}_{r,s}^{(p)}(\mathbf{x})
    =
    W_{r,s}^{(p)}\,\widetilde{\mathbf{z}}_{g}^{(p)}(\mathbf{x})
    + \mathbf{b}_{r,s}^{(p)},
\end{equation}
where $\mathbf{y}_{r,s}(\mathbf{x})$ denotes the observed voxel responses for ROI $r$ in subject $s$, $W_{r,s}^{(p)}$ is a linear mapping, and $\mathbf{b}_{r,s}^{(p)}$ is a bias term. We estimate $W_{r,s}^{(p)}$ using ridge regression on the training split,
\begin{equation}
\label{eq:ridge_objective}
    \min_{W_{r,s}^{(p)},\,\mathbf{b}_{r,s}^{(p)}}
    \sum_{\mathbf{x} \in \mathcal{D}_{\mathrm{train}}}
    \left\|
    \mathbf{y}_{r,s}(\mathbf{x}) - \widehat{\mathbf{y}}_{r,s}^{(p)}(\mathbf{x})
    \right\|_2^2
    +
    \alpha \left\|W_{r,s}^{(p)}\right\|_F^2,
\end{equation}
where the regularization parameter $\alpha$ is selected by cross-validation on the training data.

We evaluate predictivity on held-out data using Pearson correlation between predicted and observed responses, averaged across voxels within each ROI. Following prior work, we additionally compute noise-ceiling-normalized scores for NSD using the released noise-ceiling estimates. ROI-level predictivity is computed separately for each subject and constituent ROI. To summarize results at the super-category level, we first average predictivity across the ROIs belonging to that super-category within each subject and then average across the eight NSD subjects. Formally, for super-category $g$ with constituent ROI set $\mathcal{R}_g$ and subject set $\mathcal{S}$, we report
\begin{equation}
\label{eq:super_roi_average}
    \mathrm{Score}_{g}^{(p)}
    =
    \frac{1}{|\mathcal{S}|}
    \sum_{s \in \mathcal{S}}
    \frac{1}{|\mathcal{R}_g|}
    \sum_{r \in \mathcal{R}_g}
    \mathrm{corr}_{r,s}^{(p)},
\end{equation}
where $\mathrm{corr}_{r,s}^{(p)}$ denotes the held-out encoding performance for ROI $r$ in subject $s$ using the top-$p\%$ model units localized for super-category $g$. This aggregation asks whether functionally localized subsets of model units consistently predict the corresponding family of cortical regions across subjects.

\subsection{Causal Interventions on Category-Selective Regions}
\label{sec:methods:causal}

To test whether the category-selective regions identified in \ourmodel are causally responsible for category-level perception, we perform targeted activation-space interventions on the units that comprise each region. Our approach is inspired by Contrastive Activation Addition \citep[CAA;][]{rimsky-etal-2024-caa}, in which a behavioral direction is constructed as the difference between mean activations on contrastive stimulus sets, and then added to or subtracted from the residual stream to steer model behavior.

\paragraph{Constructing contrastive activation vectors.} For each category $c \in \{\text{faces, bodies, scenes, objects, words}\}$, we compute a contrastive activation vector
\[
\mathbf{v}_c = \frac{1}{|\mathcal{S}_c|}\sum_{x \in \mathcal{S}_c} \mathbf{a}(x) \;-\; \frac{1}{|\mathcal{S}_{\neg c}|}\sum_{x \in \mathcal{S}_{\neg c}} \mathbf{a}(x),
\]
where $\mathcal{S}_c$ is the set of localizer stimuli for the target category $c$, $\mathcal{S}_{\neg c}$ is the union of localizer stimuli for all other categories, and $\mathbf{a}(x)$ denotes the activation of the targeted units on stimulus $x$. Intuitively, $\mathbf{v}_c$ points from the average non-$c$ response toward the average $c$ response in activation space, isolating the direction along which the model represents category $c$.

\paragraph{Driving and suppressing.} Given a target category $c$, we apply the intervention by adding $\lambda\,\mathbf{v}_c$ to the activations of a localized set of units during the forward pass. The sign and magnitude of $\lambda$ determine the type of intervention:
\begin{itemize}
    \item \textbf{Driving} ($\lambda > 0$): adding $+\mathbf{v}_c$ pushes the activations of the targeted units toward the category-$c$ representation, biasing the model's perception toward $c$ regardless of the actual input stimulus.
    \item \textbf{Suppression} ($\lambda < 0$): adding $-\mathbf{v}_c$ pushes the activations away from the category-$c$ representation, ablating the model's ability to perceive category $c$ while leaving other categories largely intact.
\end{itemize}

\paragraph{Targeted units.} The crucial property that makes these interventions clean is that the targeted units are spatially localized on the cortical sheet. For each category, we select the top 10\% of units within the corresponding category-selective region identified by the functional localizer, and apply the intervention only to those units. Because the topographic objective concentrates each category's selective units in a compact patch of the sheet, this targeting is well-defined and does not entangle units belonging to other categories. For the driving experiment in Fig.~\ref{fig:causal-intervention}c, we vary the fraction of targeted units from 5\% to 30\% to characterize how perception scales with intervention coverage. For Fig.~\ref{fig:causal-intervention} we measure performance on a held-out set of stimuli for each category.

\subsection{Data-Driven Cluster Discovery}
\label{sec:cluster-discovery}

\subsubsection{Spacetop naturalistic movie fMRI dataset}
\label{sec:methods:fMRI_Jung_2025}

We analyzed publicly available fMRI data from the Spacetop dataset \citep{Jung2025}, including only those 83 participants who completed all 13 naturalistic movie-viewing runs. Participants watched short naturalistic video clips spanning diverse semantic content, including social interactions, nature, sports, music, and emotional narratives, while undergoing whole-brain fMRI acquisition. We used model-derived semantic clusters over 2-second video segments to define cluster-level contrasts, testing whether human cortical responses distinguished each model-predicted cluster from all other clusters. Full preprocessing, surface projection, GLM specification, contrast construction, statistical thresholding, and differences from the original \citetn{Jung2025} analysis are described below in SI~\S\ref{app:Jung}.

\subsubsection{Model-Guided Discovery via Hierarchical Clustering}
\label{subsubsec:hierarchical_clustering}

To identify functionally coherent groups of stimuli without relying on predefined category labels, we apply agglomerative hierarchical clustering to embeddings of video clips drawn from the Spacetop dataset.

\paragraph{Stimulus embeddings.}
We segment each video at 2,s intervals corresponding to the fMRI repetition time (TR) and obtain a separate embedding per TR by feeding the video up to that mark into \modelname{omni-embed-nemotron-3b} \citep{Xu2025OmniEmbedNemotronAU} and reading out the last-layer activation of the final token. We use \modelname{omni-embed-nemotron-3b} rather than \modelname{Qwen2.5-Omni-3B} to obtain embeddings that are optimised for semantic similarity rather than generation, ensuring that cluster assignments reflect conceptual content rather than next-token predictability. These per-TR embeddings serve as the input features for clustering.

\paragraph{Cortical sheet activations.}
We extract several activation maps from \ourmodel{}'s cortical sheet using the same TR-aligned segmentation: for each 2\,s mark, the model is fed the video up to that point, yielding one cortical sheet per TR. Unlike the embedding extraction, we average the per-TR sheets across the full video to obtain a single mean cortical sheet $a_i$ representing each clip.

\paragraph{Hierarchical clustering with selectivity-based stopping.}
The video clips are grouped using Ward linkage on Euclidean distances over their embeddings, producing a full binary dendrogram. We then traverse the dendrogram top-down and decide at each internal node whether to accept a candidate split. For a node containing a set of stimuli $S$, we contrast the cortical sheet activation maps of $S$ against those of the complementary set $\bar{S}$ of all remaining stimuli, computing a Welch's $t$-test independently at every cortical unit. The cluster's score is defined as the median $t$-value across units, providing a robust summary of how strongly $S$ drives a coherent population relative to the rest of the stimulus set:

\begin{equation}
\mathrm{score}(S) \;=\; \mathrm{median}_{u}\, t_{u}(S, \bar{S}),
\end{equation}

where $t_u$ denotes the unit-wise $t$-statistic. To avoid degenerate statistics and uninformative partitions, clusters with fewer than $N_{\min}=10$ or more than $N_{\max}=500$ stimuli are assigned a score of $-\infty$, preventing the recursion from accepting them as terminal nodes.

Starting from the root, we score the parent node and both candidate children at each split. If both children attain a higher score than the parent, the split is accepted and the procedure recurses into both subtrees. If both children score lower, recursion halts and $S$ is returned as a terminal cluster. If exactly one child improves on the parent, the procedure recurses into the improving subtree only and emits the other child as a terminal cluster. This early-stopping criterion retains subdivisions only when they yield more selective sub-populations, while still allowing the algorithm to descend asymmetrically into branches of the dendrogram that exhibit heterogeneous selectivity. The procedure yields a flat partition of the stimulus set in which every cluster is locally maximal under the selectivity score, subject to the size constraints. The full procedure is summarized in Algorithm~\ref{alg:cluster-discovery}.

\begin{algorithm}[t]
\caption{Top-down dendrogram traversal with selectivity-based early stopping}
\label{alg:cluster-discovery}
\begin{algorithmic}[1]
\Require Stimulus embeddings $\{x_i\}_{i=1}^{n}$, mean cortical sheet activation maps
$\{a_i\}_{i=1}^{n}$, size bounds $N_{\min}, N_{\max}$
\State $Z \gets \textsc{WardLinkage}(\{x_i\})$ \Comment{full binary dendrogram}
\State \textbf{return} $\textsc{Split}(\text{root}(Z))$
\Statex
\Function{Score}{$S$}
  \If{$|S| < N_{\min}$ \textbf{or} $|S| > N_{\max}$} \Return $-\infty$ \EndIf
  \State $\bar{S} \gets \{1,\dots,n\} \setminus S$
  \State $t_u \gets \textsc{WelchTTest}(\{a_i\}_{i \in S}, \{a_i\}_{i \in \bar{S}})$ for each unit $u$
  \State \Return $\mathrm{median}_{u}\, t_u$
\EndFunction
\Statex
\Function{Split}{$v$}
  \If{$v$ is a leaf} \Return $\{\textsc{Leaves}(v)\}$ \EndIf
  \State $L, R \gets$ children of $v$
  \State $s_p \gets \textsc{Score}(\textsc{Leaves}(v))$
  \State $s_L \gets \textsc{Score}(\textsc{Leaves}(L))$
  \State $s_R \gets \textsc{Score}(\textsc{Leaves}(R))$
  \If{$s_L < s_p$ \textbf{and} $s_R < s_p$}
    \State \Return $\{\textsc{Leaves}(v)\}$ \Comment{stop: neither child improves}
  \ElsIf{$s_L < s_p$}
    \State \Return $\{\textsc{Leaves}(L)\} \cup \textsc{Split}(R)$ \Comment{descend right only}
  \ElsIf{$s_R < s_p$}
    \State \Return $\textsc{Split}(L) \cup \{\textsc{Leaves}(R)\}$ \Comment{descend left only}
  \Else
    \State \Return $\textsc{Split}(L) \cup \textsc{Split}(R)$ \Comment{descend both}
  \EndIf
\EndFunction
\end{algorithmic}
\end{algorithm}

\subsection{Topographic ANN receptive-field mapping.}
\label{subsec:receptive-field}

To test whether \ourmodel\ develops spatially organised visual-field representations analogous to the retinotopic maps found in human visual cortex, we adapted population receptive-field (pRF) mapping methods from human fMRI to characterize visual-field preferences in \ourmodel. Classical retinotopic mapping uses rotating wedges and expanding or contracting annuli to estimate polar-angle and eccentricity preferences across cortex. pRF mapping formalizes this approach by fitting a spatial receptive-field model to each voxel's response time course, yielding estimates of preferred visual-field location and receptive-field size for the neural population sampled by that voxel \citep{wandell_visual_2007,dumoulin_population_2008}. We based our stimuli on the analyzePRF stimulus set, which combines retinotopic aperture masks with provided object/pink-noise pattern images designed to drive both low- and higher-level visual areas \citep{kay_compressive_2013,benson_human_2018}.

Unlike fMRI voxels, model units are directly observable and do not require deconvolution of a hemodynamic response. We therefore used a simpler unit-level analogue of pRF mapping. For each aperture condition \(a\), we generated 15 images \(I_{a,p}\) by applying the corresponding wedge or annulus mask to different provided pattern images \(p\), while replacing non-aperture regions with a uniform gray background.

Due to the spatial sampling of fMRI, voxels contain the aggregated response of a large population of neurons \citep{kriegeskorte_how_2010}. To simulate this readout process, we smoothed model activations with a Gaussian kernel prior to all subsequent analyses, using a unit distance of 1.0\,mm and FWHM of 4.0\,mm. Model responses were then averaged across pattern instantiations,
\[
\bar{r}_{i,a} = \frac{1}{P}\sum_{p=1}^{P} r_i(I_{a,p}),
\]
where \(r_i(I_{a,p})\) denotes the smoothed response of unit \(i\) to image \(I_{a,p}\). Polar-angle preference was assigned as
\[
\theta_i^\ast = \arg\max_{\theta} \bar{r}_{i,\theta},
\]
and eccentricity preference as
\[
e_i^\ast = \arg\max_{e} \bar{r}_{i,e}.
\]

To identify units with reliable spatial tuning, we tested each unit for a significant effect of aperture condition on its responses using a one-way ANOVA across conditions, with the $P$ patterns serving as replicates within each condition. Resulting $p$-values were corrected for multiple comparisons across units using the Benjamini--Hochberg false discovery rate procedure \citep{benjamini_controlling_1995} at a threshold of $q < 0.05$. Only units passing this criterion were classified as spatially tuned and included in the topographic analysis. The resulting unit-wise polar-angle and eccentricity estimates were plotted on the model's two-dimensional sheet to test whether visual-field preferences vary smoothly across model space.

\section*{Acknowledgements}

We thank the EPFL NeuroAI and NLP labs for useful discussions. 
M.S., B.A., and J.M. were supported by the Schmidt Science Foundation's AI2050 program.
A.G. and J.M. were supported by the Swiss National Science Foundation.
L.M. and A.A. were hosted by the Summer@EPFL program.

\section*{Code and Data Availability}

We open-source \ourmodel{}, analysis code, and pointers to data here: 
\begin{itemize}
    \item Code: \href{https://github.com/epflneuroailab/topo-omni}{github.com/epflneuroailab/topo-omni}
    \item Weights: \href{https://huggingface.co/epfl-neuroai/topo-omni}{huggingface.co/epfl-neuroai/topo-omni}
\end{itemize}
\bibliographystyle{apalike}
\bibliography{sn-bibliography}

\begin{appendices}

\section{fMRI data processing: Vision, Audio, Higher-level Cognition}
\label{app:marvi}

\paragraph{fMRI Dataset and Participants.} We analyzed all publicly available fMRI data from \citetn{Marvi2025}, using the 6 participants (nr. 1, 6, 7, 8, 9, and 21) who completed the Efficient Multifunction fMRI Localizer experiment (EMFL). The EMFL comprises 5 runs of approximately 3 minutes each (total $\sim$14 minutes scan time), in which participants viewed video stimuli drawn from 5 visual categories (faces, bodies, scenes, objects, words-on-scrambled-background) while simultaneously listening to auditory stimuli from 5 categories (false-belief stories, false-photo stories, nonwords, quilted speech, arithmetic problems). Stimuli were presented in a blocked design with a repetition time (TR) of 2 sec.

Crucially, the visual and auditory streams are assigned to blocks independently of each other, making their responses non-congruent with regard to input modality, but statistically separable within a single GLM using adequate contrasts. Because several contrasts target multiple anatomically distinct regions (e.g., Faces vs. Objects localizes FFA, OFA, and fSTS simultaneously), this orthogonal design allows up to 14 functional regions spanning visual, language, theory-of-mind, speech, and multiple-demand networks to be localized from 9 contrasts in about 14 minutes scanning time per subject - roughly one third of the time a conventional localizer battery would require \citep{Marvi2025}.

\paragraph{Pre-processing.} We pre-processed the raw BIDS-formatted data using fMRIprep 24.0.1 \citep{esteban_fmriprep_2019}, and we used FreeSurfer 7.3.2 for cortical surface reconstruction. We applied all default fMRIprep preprocessing steps, including slice-timing correction, head motion estimation, susceptibility distortion correction, and co-registration to the T1w image. For the main fROI analysis (replicating Figure 4 in \citetn{Marvi2025}), we used BOLD data projected to MNI volumetric space at 2 mm isotropic resolution, following \citetn{Marvi2025}. For cortical surface visualization (replicating Figures 2 and 3 in \citetn{Marvi2025}), we used BOLD data in native T1w volumetric space, from which we projected statistical maps onto each subject's individual FreeSurfer native cortical surface (fsnative).

\paragraph{First-Level general linear model.} We estimated subject-level general linear models (GLMs) using Nilearn 0.12.1 \citep{abraham_machine_2014}. Each GLM design matrix included one regressor per stimulus condition (10 conditions total: 5 visual, 5 auditory/cognitive), modeled by convolution with the canonical hemodynamic response function. As nuisance regressors, we included 6 rigid-body head motion parameters (3 translations, 3 rotations). Additionally, we included a first-order polynomial drift term to account for low-frequency signal trends (equivalent to a high-pass filter with cutoff of 0.01 Hz) and assumed an AR(1) autoregressive noise model. We applied spatial smoothing with a 3 mm FWHM Gaussian kernel, matching the spatial smoothing used in the original pipeline used in \citetn{Marvi2025}.

We fit GLMs separately for each run and each subject. For cross-validated fROI analyses, we additionally fit GLMs on two run subsets: even runs (runs 2 and 4) and odd runs (runs 1, 3, and 5).

We computed the following nine EMFL contrasts, matching the contrasts reported in Table 3 in 
\citetn{Marvi2025}:

\begin{table}[h]
\centering
\begin{tabular}{lll}
\toprule
\textbf{Contrast} & \textbf{Formula} & \textbf{Target ROI} \\
\midrule
Faces $>$ Objects 
& faces $-$ objects 
& FFA, OFA, fSTS \\

Scenes $>$ Objects 
& scenes $-$ objects 
& PPA, OPA, RSC \\

Bodies $>$ Objects 
& bodies $-$ objects 
& EBA \\

Words $>$ Objects 
& words\_scr\_objects $-$ objects 
& VWFA \\

Objects $>$ Words 
& objects $-$ words\_scr\_objects 
& LOC \\

False Belief $>$ False Photo 
& false\_belief $-$ false\_photo 
& rTPJ \\

English $>$ Nonwords 
& english\_stories $-$ nonwords 
& Language network \\

Nonwords $>$ Quilted 
& nonwords $-$ quilted\_speech 
& Speech-selective STG \\

Math $>$ English Stories 
& math $-$ (false\_belief $+$ false\_photo) / 2 
& Multiple Demand network \\
\bottomrule
\end{tabular}
\caption{Overview of EMFL contrasts and targeted regions of interest. The contrasts follow those reported in \citetn{Marvi2025}.}
\label{tab:contrasts}
\end{table}

\paragraph{Functional ROI Definition and Cross-Validated Response Extraction (replicating Figure 4 in \citealp{Marvi2025}).} To replicate the functional ROI (fROI) analysis of Figure 4 in \citetn{Marvi2025}  we followed their methods. We downloaded anatomical constraint parcels from the \href{https://github.com/aimarvi/emfl_analysis}{EMFL GitHub repository}, using the same parcel files as \citetn{Marvi2025}.

Within each anatomical parcel, we defined an fROI as the top 10\% of voxels by t-statistic for the relevant functional contrast, computed using the GLM fit on a held-out run split, thus following \citetn{Marvi2025}. We then extracted responses from the resulting fROI using the complementary run split (cross-validation). Specifically, following \citetn{Marvi2025}:

\begin{itemize}
    \item \textbf{Split A:} We defined the fROI using even runs (2, 4) and extracted condition responses using odd runs (1, 3, 5).
    \item \textbf{Split B:} We defined the fROI using odd runs (1, 3, 5) and extracted condition responses using even runs (2, 4).
\end{itemize}

For each of the 10 conditions, we extracted mean beta estimates by averaging within the fROI mask across voxels and across held-out runs, then averaged across both cross-validation splits (A and B), and finally averaged across subjects. We display results as group mean ± SEM with individual subject data overlaid, directly replicating the format of Figure 4 in \citetn{Marvi2025}. The original study used 20 subjects, whereas our analysis uses the 6-subject subset available in the public \href{https://doi.org/10.18112/openneuro.ds006179.v1.0.1}{OpenNeuro release}.

We find that our results closely match those in 
Figure 4 in \citetn{Marvi2025}. For example, left FFA shows strong selectivity for faces versus all other conditions, consistent with the original paper.

\paragraph{Surface Visualization (replicating Figures 2 and 3 in \citep{Marvi2025}).} To generate individual-subject cortical surface maps comparable to Figures 2 and 3 in \citetn{Marvi2025}, we estimated a concatenated first-level GLM on BOLD data in native T1w space, pooling all 5 runs into a single design matrix (concatenated in time) with the same GLM parameters described above. We projected statistical maps from native T1w volumetric space onto each subject's FreeSurfer native cortical surface, consistent with \citetn{Marvi2025}. We display activation maps as signed log $p$-values ($-\log_{10}(p) \times \text{sign}(t)$), thresholded at $\pm 3$ (equivalent to $p < 0.001$, uncorrected), on inflated cortical surfaces and overlay ROI contours from independent studies:

\begin{itemize}
    \item \textbf{Visual ROIs} (FFA, OFA, fSTS, PPA, OPA, RSC, EBA, LOC): \citetn{Julian2012} parcels in CVS template space
    \item \textbf{VWFA}: \citetn{Saygin2016} parcel in CVS-MNI152 template space
    \item \textbf{Theory of Mind} (rTPJ and medial prefrontal regions): \citetn{Dufour2013} parcels in MNI152 space
    \item \textbf{Language network} (IFG, IFGorb, MFG, AntTemp, PostTemp, AG): \citetn{Fedorenko2010} parcels in MNI152 space
    \item \textbf{Speech} (bilateral STG): \citetn{Regev2025} parcels in MNI152 space
    \item \textbf{Multiple Demand network} (frontal and parietal regions): \citetn{Fedorenko2013} parcels in MNI152 space
\end{itemize}

\paragraph{Differences from \citep{Marvi2025}:} The original paper used FS-FAST and FreeSurfer for all preprocessing and GLM estimation, operating throughout in native FreeSurfer subject space. We used fMRIprep for preprocessing and Nilearn for GLM estimation. The key functional analyses (parcel source, fROI definition, cross-validation) are identical, whereas some preprocessing steps and coordinate spaces differ.

\begin{figure}[H] 
\centering
\includegraphics[width=1\textwidth]{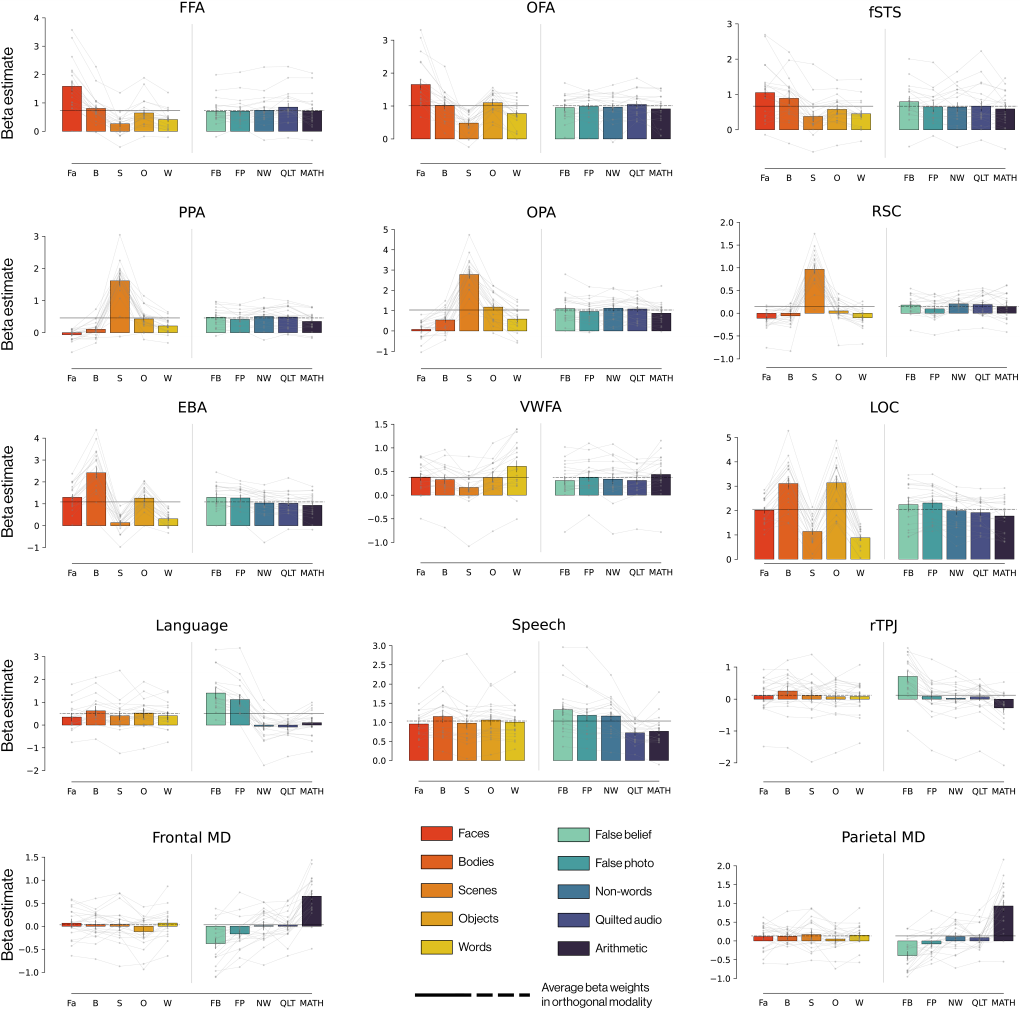}
\caption{
        \textbf{Localizer results: original analysis from Figure 4 in \citetn{Marvi2025} based on 20 subjects.} Figure copied from \citetn{Marvi2025} under the Creative Commons Attribution 4.0 International (CC BY 4.0) license. For the results of our re-analysis based on 6 subjects for which data are publicly available, see Fig.~\ref{fig:apx:marvi_omni_reanalysis}.
}
\label{fig:apx:marvi_original_analysis}
\end{figure}

\begin{figure}[H] 
\centering
\includegraphics[width=1\textwidth]{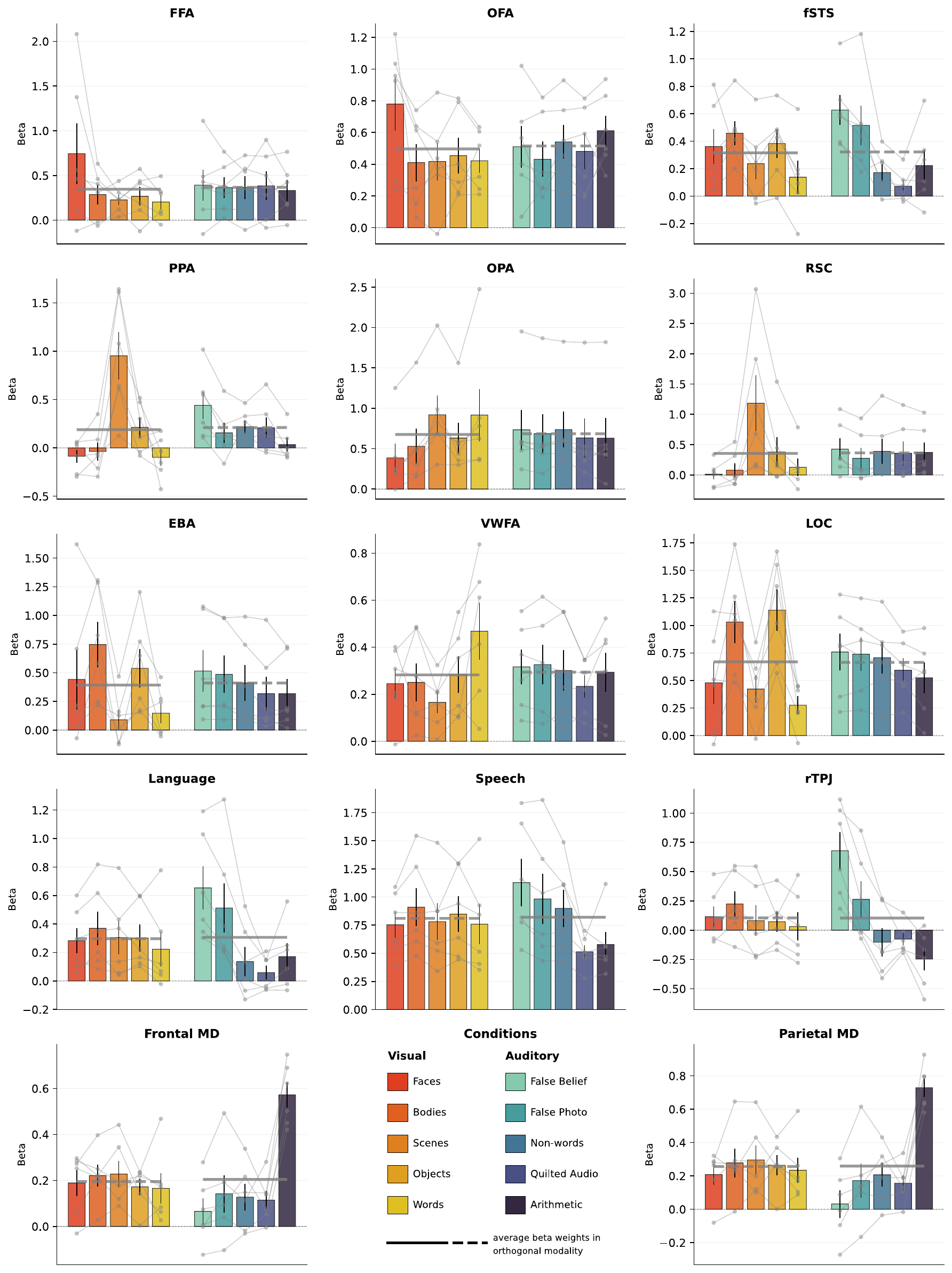}
\caption{
        \textbf{Localizer results: re-analysis of data from \citetn{Marvi2025} based on 6 publicly available subjects.} For the original results based on 20 subjects, please see Fig.~\ref{fig:apx:marvi_original_analysis}
}
\label{fig:apx:marvi_omni_reanalysis}
\end{figure}

\section{fMRI data processing: Human Voice-Selective Areas}
\label{app:pernet}

\paragraph{fMRI Dataset and Participants.} We analyzed publicly available fMRI data from all 218 participants who completed the experiment described in \citetn{Pernet2015}. The data are available from the \href{https://datashare.ed.ac.uk/handle/10283/818}{Edinburgh DataShare repository}). The paradigm followed a block design in which participants passively listened to 20 vocal sounds and 20 non-vocal sounds (8 s each), interleaved with 20 blocks of silence (8 s each), for a total acquisition duration of 620 s (310 volumes), whereby stimulus presentation order was fixed and identical for all 218 participants. 

Vocal stimuli contained sounds of human vocal origin from 47 speakers — including speech (words, syllables, and sentence fragments in English, French, Finnish, and Arabic) and non-speech sounds (e.g., laughs, cries, coughs) — produced by speakers spanning a wide age range (infants to elderly adults). Non-vocal stimuli contained natural environmental sounds (e.g., water, wind, animal calls) and man-made sounds (e.g., vehicles, instruments, classical music excerpts). All stimuli were 16-bit mono audio recorded at a sampling rate of 22,050 Hz and normalized to the same root-mean-square (RMS) amplitude, so that all sounds had equal average acoustic energy regardless of category. Stimuli were presented via MRI-compatible headphones at 80–85 dB. 

\paragraph{Pre-processing.} We preprocessed the raw fMRI data using FSL 6.0.7 \citep{jenkinson_fsl_2012} following the volumetric analysis pipeline described in \citetn{Pernet2015}, which was originally implemented in SPM12b. For each participant, we applied the following steps in sequence: (1) slice timing correction, (2) motion correction using a 6-degree-of-freedom (DOF) rigid-body model, (3) coregistration of the T1-weighted anatomical image to the mean functional image, (4) nonlinear normalization to MNI152 2 mm isotropic space, (5) spatial smoothing with a 6 mm FWHM isotropic Gaussian kernel.

\paragraph{First-Level general linear model.} We estimated subject-level GLMs using Nilearn 0.10.4 \citep{abraham_machine_2014}. The design matrix for each participant included three task regressors — vocal, non-vocal, and silence — modeled by convolving a boxcar function for each block with the canonical hemodynamic response function (SPM double-gamma parameterization as implemented in Nilearn). Temporal drift was modeled using a cosine basis set (high-pass cutoff: 128 s). We assumed an AR(1) autoregressive noise model.

As nuisance regressors, we included 6 rigid-body head motion parameters (3 translations, 3 rotations) extracted from the SPM realignment transformation matrices provided with the original \citetn{Pernet2015} dataset, as well as their temporal derivatives (6 additional regressors). We additionally included one spike regressor per identified outlier volume as surpassing a certain level of framewise displacement following \citetn{carling_resistant_2000}. 

We computed the contrast vocal vs. non-vocal for each participant, with TRs of the silence condition assigned a weight of zero.

\paragraph{Group-Level Analysis.} We entered the 218 individual-level contrast images (vocal $>$ non-vocal) into a one-sample t-test using Nilearn's second-level GLM, implementing the random-effects analysis described in \citetn{Pernet2015} Figure 2. We applied family-wise error (FWE) correction via Gaussian random field theory at p < 0.05, corresponding to a t-threshold of 1.96, with a minimum cluster extent of 10 voxels. All group-level maps are in MNI152 2 mm isotropic space (99 × 117 × 95 voxels).

\paragraph{Surface Visualization.} For display purposes, we projected the group-level t-statistic map from MNI152 volumetric space to the fsaverage6 cortical surface template ($\sim$82,000 vertices; 40,962 per hemisphere). We display only the FWE-thresholded positive t-values (vocal vs. non-vocal selective regions), matching the visualization format of \citetn{Pernet2015} Figure 2.

\begin{figure}[H] 
\centering
\includegraphics[width=1\textwidth]{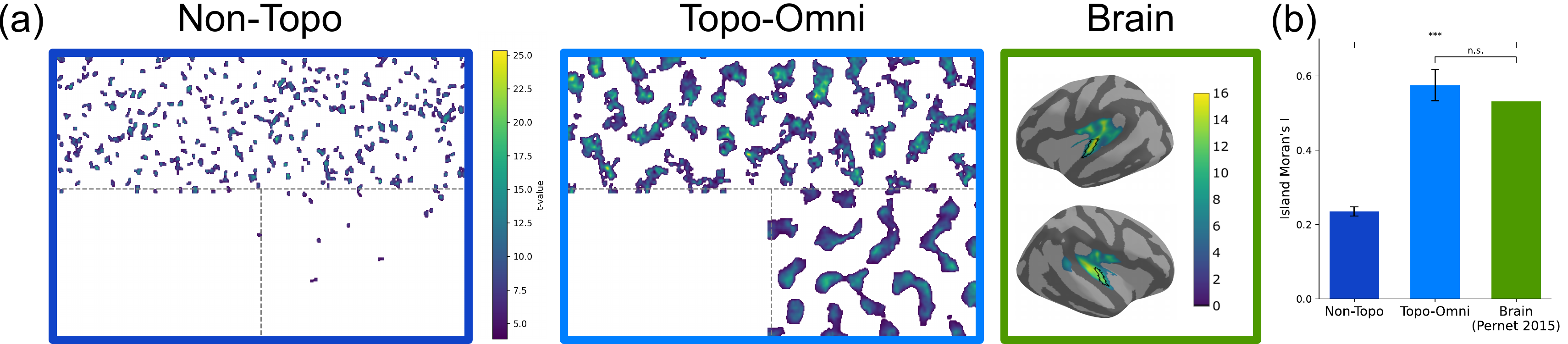}
\caption{\textbf{Clustering of vocal vs. non-vocal responses in model and brain. A)} Contrasting vocal and non-vocal stimuli from \citet{Pernet2015} yields a network of large human-voice selective clusters in \ourmodel as shown in the superior temporal sulcus (center and right panel). In comparison, the non-topographic counterpart of \ourmodel shows relatively small selectivity clusters that mainly arise due to the fwhm-smoothing (for details, see \ref{app:topo-clustering}) we applied to simulate the fMRI-readout process (left panel). \textbf{B) } We quantified the clustering patterns using the spatial auto-correlation metric Island Moran's I (
for details, see \citet{rathi_topolm_2025}) that reflects clustering of islands - of units in a model sheet or of vertices representing the cortical sheet - that show a significant contrast response after correction for multiple comparison (here, using FDR at $q < 0.05$). At the level of Island Moran's I \ourmodel is indistinguishable from the brain, whereas the non-topographic counterpart of \ourmodel shows a significantly reduced level of clustering. }
\label{fig:apx:pernet}
\end{figure}

\paragraph{Clustering analysis.} To quantify the clustering of the vocal-selective activation pattern, we computed island Moran's I on the group-level vocal $>$ non-vocal t-map projected to the fsaverage6 surface. We identified contiguous clusters of FDR-significant vertices ($q < 0.05$, minimum island size: 8 vertices) and computed Moran's I within each cluster. The mean island Moran's I across both hemispheres was then compared against the per-island distributions from \ourmodel and its non-topographic counterpart (Figure~\ref{fig:apx:pernet}). In terms of Island Moran's I, \ourmodel is indistinguishable from the brain activation patterns found in \citet{Pernet2015} (one-sample $t$-test: $t(78) = 1.04$, $p = .850$; Wilcoxon signed-rank: $p = .867$), whereas its non-topographic counterpart displays significantly lower clustering ($t(417) = -24.61$, $p < .001$; Wilcoxon signed-rank: $p < .001$). 

For comparisons of the degree of clustering in response to  other contrasts (in other modalities), see SI \S\ref{app:topo-clustering}.

\paragraph{Cross-validated temporal-voice-acrea  response profile.} To obtain unbiased estimates of vocal and non-vocal response magnitudes within the temporal voice areas (TVA), we performed a cross-validated functional region-of-interest (fROI) analysis following \citetn{Marvi2025}. Because \citetn{Pernet2015} comprises a single run per participant, we adapted the odd/even run split of \citetn{Marvi2025} to a block-level split within the single run: the 20 vocal and 20 non-vocal blocks were each randomly partitioned into two sets of 10, yielding fold A and fold B.
\label{apx:pernet_cross_validated_response_profile}

For each participant, two first-level GLMs were estimated on the same preprocessed time series - one including only the fold-A blocks, one including only the fold-B blocks - with all other design matrix components (motion regressors, drift basis) held constant. Group-level one-sample $t$-tests were then performed separately on the 218 fold-A contrast images and the 218 fold-B contrast images, using the same
second-level GLM and FWE correction procedure as described above. The resulting FWE-thresholded maps defined the fold-A and fold-B fROI masks.

Per-participant response estimates were extracted cross-validated: mean vocal and non-vocal betas within the fold-A fROI were taken from the held-out fold-B GLM, and vice versa. Following \citetn{Marvi2025}, the final response estimate for each condition was the average across the two cross-validated estimates, $\hat{\beta} = (\hat{\beta}_{\text{fold A}} + \hat{\beta}_{\text{fold B}}) / 2$. Group means $\pm$ SEM across all 218 participants are reported as a two-bar profile (vocal, non-vocal) in Fig.~\ref{fig:audio-clusters}.

\section{fMRI data: Tonotopic organization in the audio encoder}
\label{app:Hedger}
The human auditory cortex is organized tonotopically: neighboring cortical locations prefer neighboring sound frequencies, yielding a smooth map of preferred frequency across the cortical surface. We tested whether a spatially organized frequency map emerges in the audio encoder of \ourmodel.

We presented pure tones spanning [100--7000]\,Hz and, following the unit-level receptive-field procedure described for retinotopy (\S\ref{subsec:receptive-field}), characterized each unit's spectral tuning.
For each frequency condition $f$ we obtained the unit's mean response across the $n$ tone exemplars presented at that frequency, and assigned each unit a
preferred frequency
\[
f_i^\ast = \arg\max_{f} \bar{r}_{i,f}.
\]

As for retinotopy, we identified reliably tuned units with a one-way ANOVA testing for an effect of frequency condition on each unit's responses (tone exemplars as replicates within each condition), correcting across units with the Benjamini--Hochberg false discovery rate procedure \citep{benjamini_controlling_1995} at $q < 0.05$. Only units passing this criterion were retained, and their preferred-frequency estimates were plotted on the model's two-dimensional sheet to test whether spectral preferences vary smoothly across model space.

We find that the audio encoder of \ourmodel\ develops spatially organized frequency preferences, with neighboring units tending to share similar best frequencies, consistent with the local tonotopic organization of human auditory cortex \citep{Hedger2026}. We emphasize that the in-silico sheet captures this local frequency smoothness rather than the single, globally ordered low-to-high gradient observed along an anatomical landmark such as Heschl's gyrus. We see this as consistent with our framing throughout, \ourmodel captures organizational principles (here, the co-localization of similarly tuned units) rather than the cortical anatomy on which the human gradient unfolds.

\section{fMRI data processing: cluster discovery}
\label{app:Jung}

\paragraph{fMRI Dataset and Participants.}
We analyzed publicly available fMRI data from 83 participants drawn from the Spacetop dataset \citep{Jung2025}, a large-scale naturalistic neuroimaging dataset in which participants watched 49 short video clips spanning a diverse range of semantic content — social interactions, nature, sports, music, and emotional narratives — while undergoing whole-brain fMRI acquisition. Each clip was presented once per participant and was followed by a structured emotion-rating epoch (~35 s) during which participants answered seven questions covering personal relevance, happiness, sadness, fear, disgust, warmth, and engagement. Videos were distributed across 13 functional runs over four sessions; we included only participants who completed all 13 task runs. (The video "tornado" appears twice in the stimulus schedule (\citetn{Jung2025}, Table 5: ses-02 run-03 and ses-04 run-02), despite the dataset description stating 49 unique videos with no repetitions. We treated the two presentations as the same stimulus identity pooling both into a single "tornado" condition in the GLM (target vs. all other videos). Overall, this yields 48 unique video identities for individual video contrasts.)

\paragraph{Pre-processing.}
We processed the raw BIDS-formatted Spacetop data using fMRIPrep 24.0.1 \citep{esteban_fmriprep_2019}, applying all default preprocessing steps, including slice-timing correction, head motion estimation, B0 fieldmap-based susceptibility distortion correction, boundary-based BOLD-to-T1w coregistration, anatomical reconstruction with FreeSurfer 7.1, and surface projection via \texttt{mri\_vol2surf}. We projected BOLD data onto the fsaverage6 surface template (40,962 vertices per hemisphere; 81,924 total) to reduce storage and computation relative to the full fsaverage surface used by \citetn{Jung2025}. Nuisance regression was applied per run using ordinary least squares with 24 confound regressors: six rigid-body motion parameters and their temporal derivatives (12 total), five anatomical and three temporal CompCor components estimated by fMRIPrep (8 total), and four cosine basis functions capturing low-frequency scanner drift. TRs flagged as motion outliers (framewise displacement $>$ 0.9 mm or standardized DVARS $>$ 1.5) were retained and their residual variance was absorbed by the nuisance model.

\paragraph{First-Level general linear model.}
We estimated subject-level GLMs using Nilearn \citep{abraham_machine_2014}. For each participant and run, stimulus epochs were modeled as boxcar functions convolved with a canonical double-gamma hemodynamic response function (SPM parameterization; peak ~5 s). T-statistics were derived from estimated contrast vectors over the OLS standard error ($\hat{\beta} = (X^\top X)^{-1} X^\top Y$, where $Y$ is the time $\times$ vertices BOLD matrix). Group-level inference used one-sample t-tests across participants (df = $n_\text{subjects} - 1 = 82$). The GLM design and confound set were identical across all contrast types; contrasts differ only in the definition of the regressor of interest.

\paragraph{Contrast Design.}
We assigned 2-second segments of all 49 videos to one of 14 semantic clusters based on the model's internal representations and using a hierarchical clustering approach (Section~\ref{subsubsec:hierarchical_clustering}). For each cluster, TRs during assigned segments were modeled as a single regressor and contrasted against TRs from all other clusters; emotion-rating TRs were left unmodeled. This contrast tests whether the brain distinguishes the semantic content of a given model-defined cluster from all other clusters. We report results for two clusters whose vertices showed significant t-values after correction for multiple comparison across the entire cortex (for details, see below): an \emph{animals} cluster (135 segments drawn from the video \emph{planetearth}) and a \emph{nature} cluster (105 segments drawn exclusively from the video \emph{mountainbike}).

\paragraph{Statistical Thresholding.}
Statistical maps were thresholded using the Benjamini-Hochberg false discovery rate (FDR) procedure \citep{benjamini_controlling_1995} at $q < 0.05$. We performed one-sided group-level t-statistics at each fsaverage6 vertex and converted them to p-values. The correction was applied jointly across both hemispheres (81,924 vertices) within each contrast independently. For visualization, maps are further restricted to the top 10\% of significant vertices by t-statistic after the correction of multiple comparison, computed jointly across both hemispheres.

\paragraph{Differences from \citep{Jung2025}.}
The original study preprocessed the same dataset using fMRIPrep 21.0.2 and reported BOLD data on the full fsaverage surface (163,842 vertices per hemisphere). We used fMRIPrep 24.0.1 and the lower-resolution fsaverage6 surface. \citetn{Jung2025} do not enumerate their GLM confound list explicitly; we used 24 regressors drawn from the standard fMRIPrep confound output. Rather than validating the dataset with a single all-videos-versus-rating-baseline contrast, we define per-cluster contrasts targeting semantic selectivity predicted by \ourmodel.

\begin{figure}[H] 
\centering
\includegraphics[width=1\textwidth]{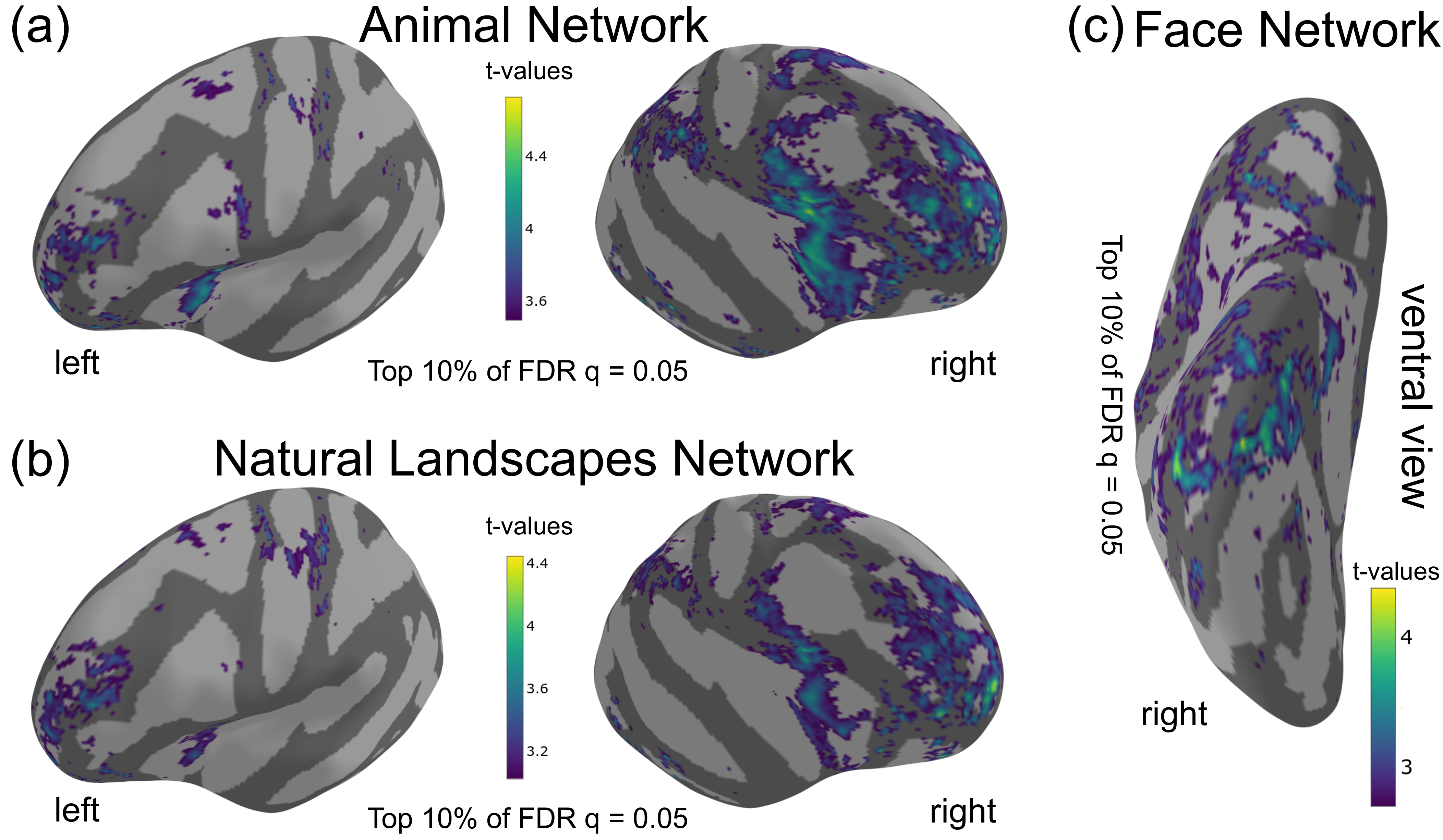}
\caption{\textbf{Model-guided discovery of 3 cortical networks.} Networks derived via the discovery pipeline (Methods Section~\ref{sec:cluster-discovery}) and validated on the Spacetop human fMRI data from \citetn{Jung2025}: (a) animals eliciting activity in frontal pole and lateral pre-frontal cortex, mostly in the right hemisphere, (b) natural landscapes similarly eliciting activity in the frontal pole and lateral pre-frontal cortex, (c) faces eliciting activity in anterior regions of the inferior temporal cortex (ventral view). Maps show the top 10\% of FDR-significant vertices (q = 0.05).
}
\label{fig:apx:cluster_exploration_netowork_brain_maps_1}
\end{figure}

\paragraph{Model-guided discovery of a face network.} 
Applying the discovery pipeline (Methods Section~\ref{sec:cluster-discovery}) to the Spacetop fMRI data \citep{Jung2025}, we combined hierarchical clustering over semantic stimulus embeddings with cortical-sheet selectivity profiles to derive candidate contrasts and tested them against human fMRI (Fig.~\ref{fig:apx:cluster_exploration_netowork_brain_maps_1}). The procedure recovered three reliable networks. A ventral face network (Fig.~\ref{fig:apx:cluster_exploration_netowork_brain_maps_1}c) served as a positive control. It is not exactly located where canonical inferior temporal cortex face regions FFA lies, likely because the model-selected images (faces in interview-like settings vs. all other videos) differ from traditional localizers. The other two networks are, to our knowledge, not described via a comparable contrast: one selective for animals (Fig.~\ref{fig:apx:cluster_exploration_netowork_brain_maps_1}a) and one for natural landscapes (Fig.~\ref{fig:apx:cluster_exploration_netowork_brain_maps_1}b). Human responses to the model-derived segments validated each prediction (top 10\% of FDR-significant vertices, q = 0.05)

\begin{figure}[H] 
\centering
\includegraphics[width=0.7\textwidth]{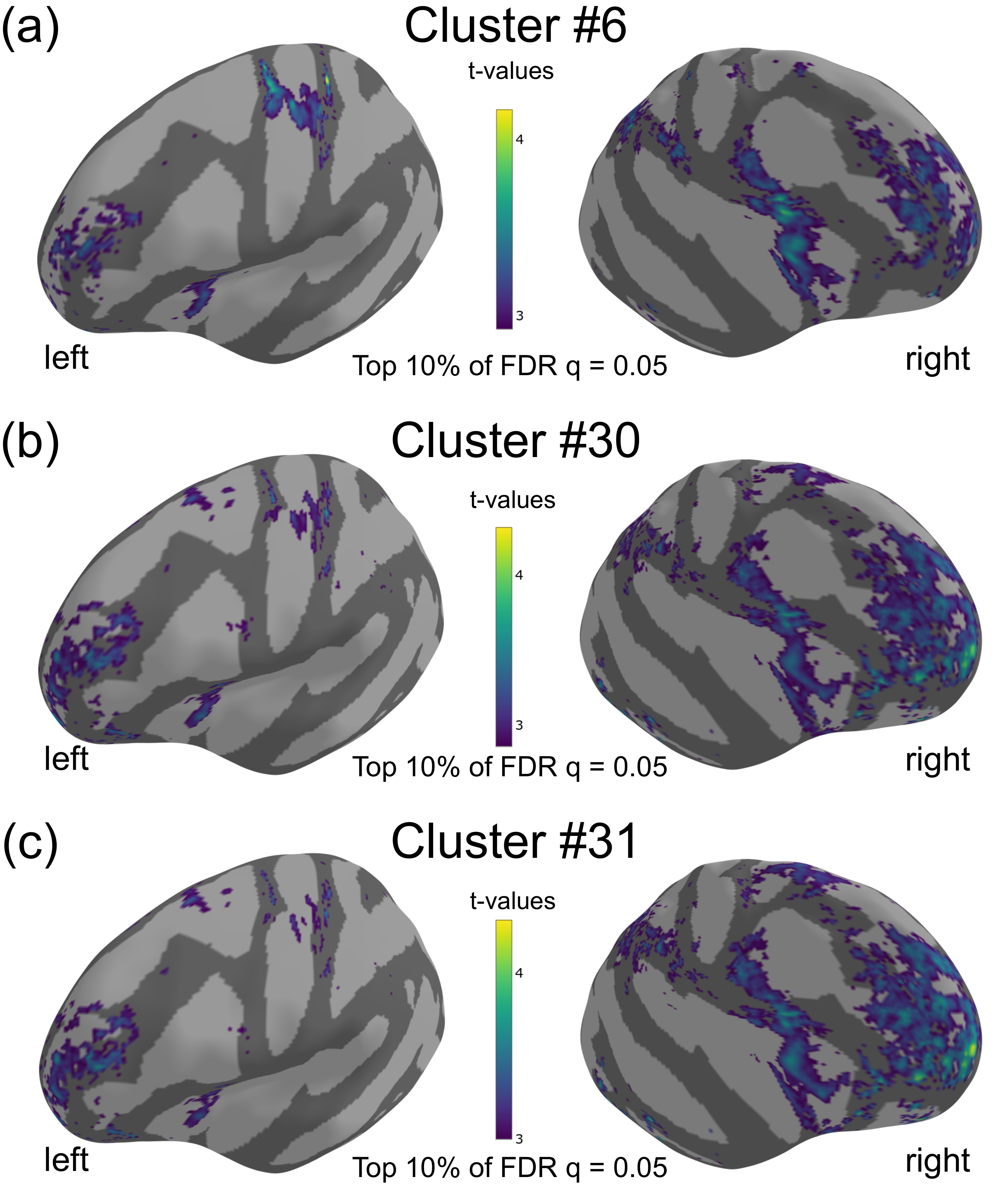}
\caption{\textbf{Model-guided discovery of additional cortical networks.} Additional networks derived via the discovery pipeline (Methods Section~\ref{sec:cluster-discovery}) and validated on the Spacetop human fMRI data from \citetn{Jung2025}: (a) cluster \#6: animals in predator and prey roles hunting each other eliciting a network largely overlapping with the animal cluster described in (Fig.~\ref{fig:apx:cluster_exploration_netowork_brain_maps_1}a), but additionally yielding responses in the left somatosensory cortex  (b) \& (c) clusters \#30 and \#31:  additional natural landscapes that parallel the natural landscapes cluster described in (Fig.~\ref{fig:apx:cluster_exploration_netowork_brain_maps_1}b) in both stimuli and cortical location. Maps show the top 10\% of FDR-significant vertices (q = 0.05).
}
\label{fig:apx:cluster_exploration_netowork_brain_maps_2}
\end{figure}

\section{Spatial clustering of selective units requires the topographic objective}
\label{app:topo-clustering}

\begin{figure}[H]
    \centering
    \includegraphics[width=1\linewidth]{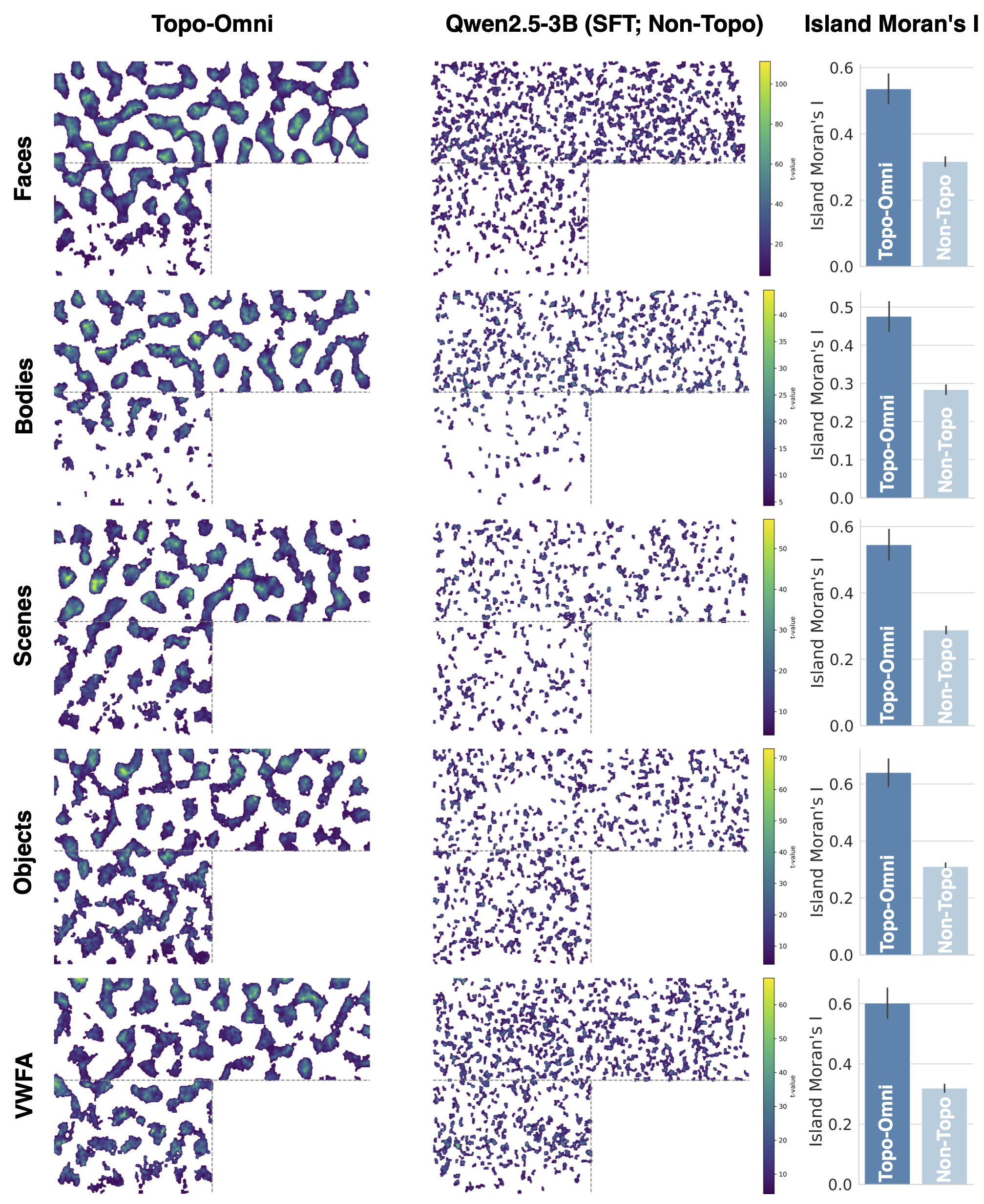}
    \caption{\textbf{Topographic training clusters visual category selectivity.} Per-unit selectivity $t$-values for five visual localizers (faces, bodies, scenes, objects, visual words/VWFA), thresholded at $p < 0.001$ (FDR-corrected) and smoothed to approximate fMRI sampling (Gaussian FWHM $= 4.0$\,mm). \textbf{Left:} \ourmodel{} (Topo) forms contiguous selective clusters. \textbf{Middle:} the non-topographic counterpart (Qwen2.5-3B SFT; Non-Topo) is sparse and salt-and-pepper. \textbf{Right:} island Moran's $I$ per localizer (higher $=$ more clustered; error bars: SEM across islands). Units are localized over the full cortical sheet, not the vision partition alone.}
    \label{fig:app-vision-comparison}
\end{figure}

\begin{figure}[H]
    \centering
    \includegraphics[width=1\linewidth]{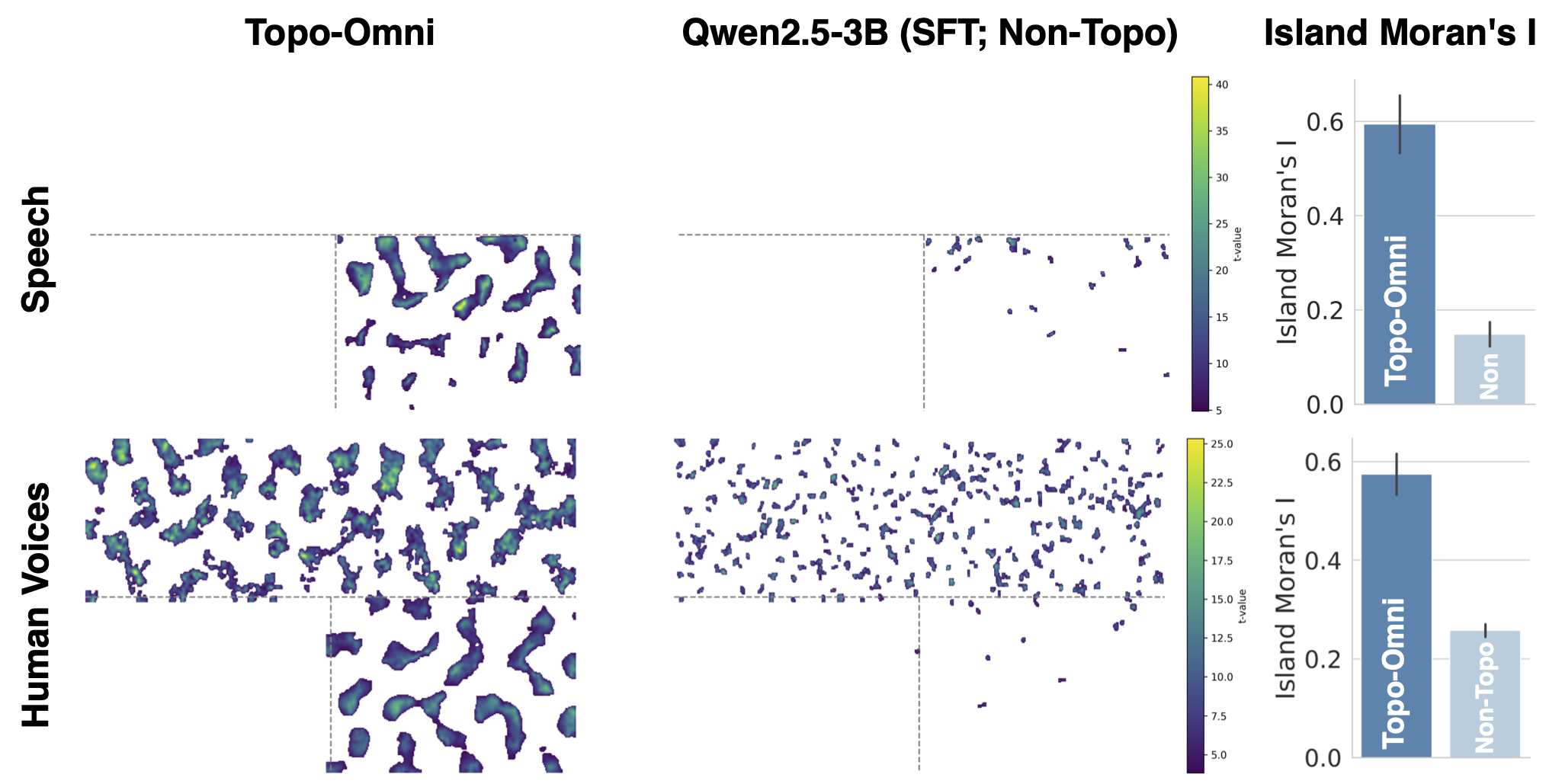}
    \caption{\textbf{Topographic training clusters auditory selectivity.} As in
    Fig.~\ref{fig:app-vision-comparison}, for the speech and human-voice localizers.
    \ourmodel{} (Topo, left) forms contiguous selective clusters, while the
    non-topographic counterpart (Non-Topo, middle) is salt-and-pepper.
    \textbf{Right:} island Moran's $I$ per localizer confirms substantially stronger
    clustering in \ourmodel{} (error bars: SEM across clusters).}
    \label{fig:app-audio-comparison}
\end{figure}

\begin{figure}[H]
    \centering
    \includegraphics[width=1\linewidth]{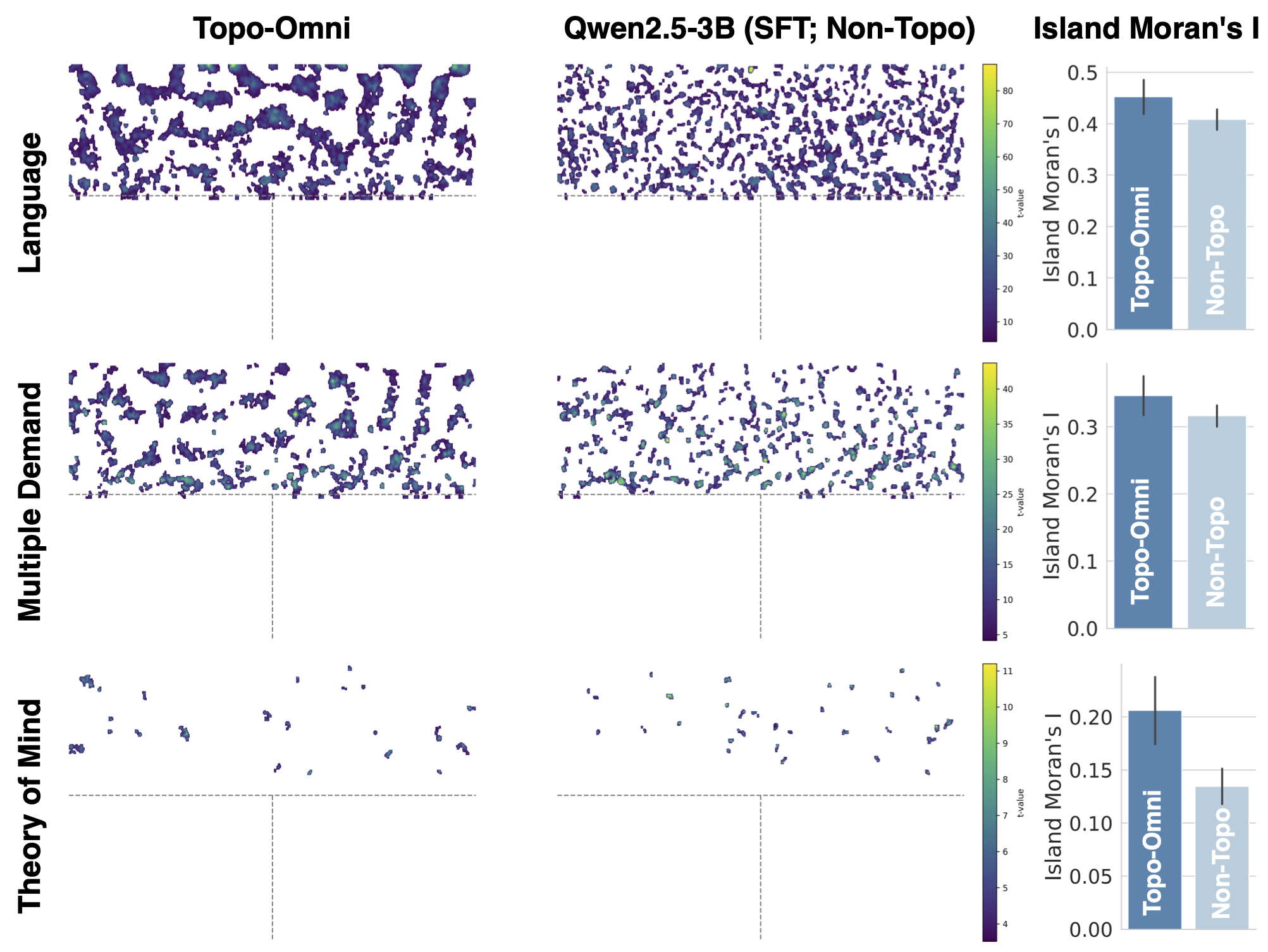}
    \caption{\textbf{Topographic training clusters higher-cognitive selectivity, with a smaller margin over the non-topographic baseline.} As in Fig.~\ref{fig:app-vision-comparison}, for the language, multiple-demand (MD), and theory-of-mind (ToM) localizers. \ourmodel{} (Topo, left) again forms more contiguous selective clusters than the non-topographic counterpart (Non-Topo, middle), but the gap in island Moran's $I$ (right) is markedly smaller than for the visual and auditory localizers. We attribute this to input modality: these localizers are driven by text tokens, whereas the cortical sheet was trained on audiovisual tokens, so $\mathcal{L}_{\text{spatial}}$ shapes the text-driven representations less directly. ToM units are thresholded at $p < 0.05$ (FDR-corrected) rather than the $p < 0.001$ used elsewhere, as no units survived the stricter threshold. Error bars: SEM across clusters.}
    \label{fig:app-cognitive-comparison}
\end{figure}

To isolate the effect of $\mathcal{L}_{\text{spatial}}$, we compare \ourmodel{} against a non-topographic counterpart (Qwen2.5-3B SFT) fine-tuned on identical data with the spatial loss disabled, and repeat this comparison across all three modeled domains: visual categories (Fig.~\ref{fig:app-vision-comparison}: faces, bodies, scenes, objects, visual words), auditory categories (Fig.~\ref{fig:app-audio-comparison}: speech, human voices), and higher-cognitive networks (Fig.~\ref{fig:app-cognitive-comparison}: language, multiple-demand, theory-of-mind). For every localizer we compute a per-unit selectivity $t$-value (Welch's $t$-test, preferred vs.\ non-preferred stimuli), threshold at $p < 0.001$ with FDR correction, and smooth the surviving map with a Gaussian kernel (FWHM $= 4.0$\,mm) to approximate the spatial sampling of fMRI; the theory-of-mind localizer is the sole exception, thresholded at $p < 0.05$ because no units survived $p < 0.001$.

Both models recover selective populations of comparable size and strength, confirming that selectivity itself is a property of the shared backbone. Their spatial layout, however, differs sharply: \ourmodel{} organizes selective units into large, contiguous clusters, whereas the non-topographic model scatters them in a salt-and-pepper pattern with no coherent structure. We quantify this with the island Moran's $I$ of each selective map (higher values indicate stronger local clustering, for imple), reported per localizer in the right column of each figure. \ourmodel{} attains higher island Moran's $I$ than its non-topographic counterpart for every localizer, establishing that the spatial organization is induced by $\mathcal{L}_{\text{spatial}}$ rather than by the training data. The margin is large for the visual and auditory localizers but substantially smaller for the higher-cognitive networks (Fig.~\ref{fig:app-cognitive-comparison}); we attribute this to those localizers being driven by text tokens, whereas the cortical sheet was trained on audiovisual tokens, leaving the text-driven representations less directly shaped by the spatial objective.

These localizers are not anatomically constrained: for each contrast we test every unit in the full cortical sheet rather than restricting the search to the corresponding modality partition. 

\section{Additional response-profile analyses}

\subsection{Bodies localizer}
\label{sec:appendix:bodies}

\begin{figure}[H]
\centering
\includegraphics[width=0.95\textwidth]{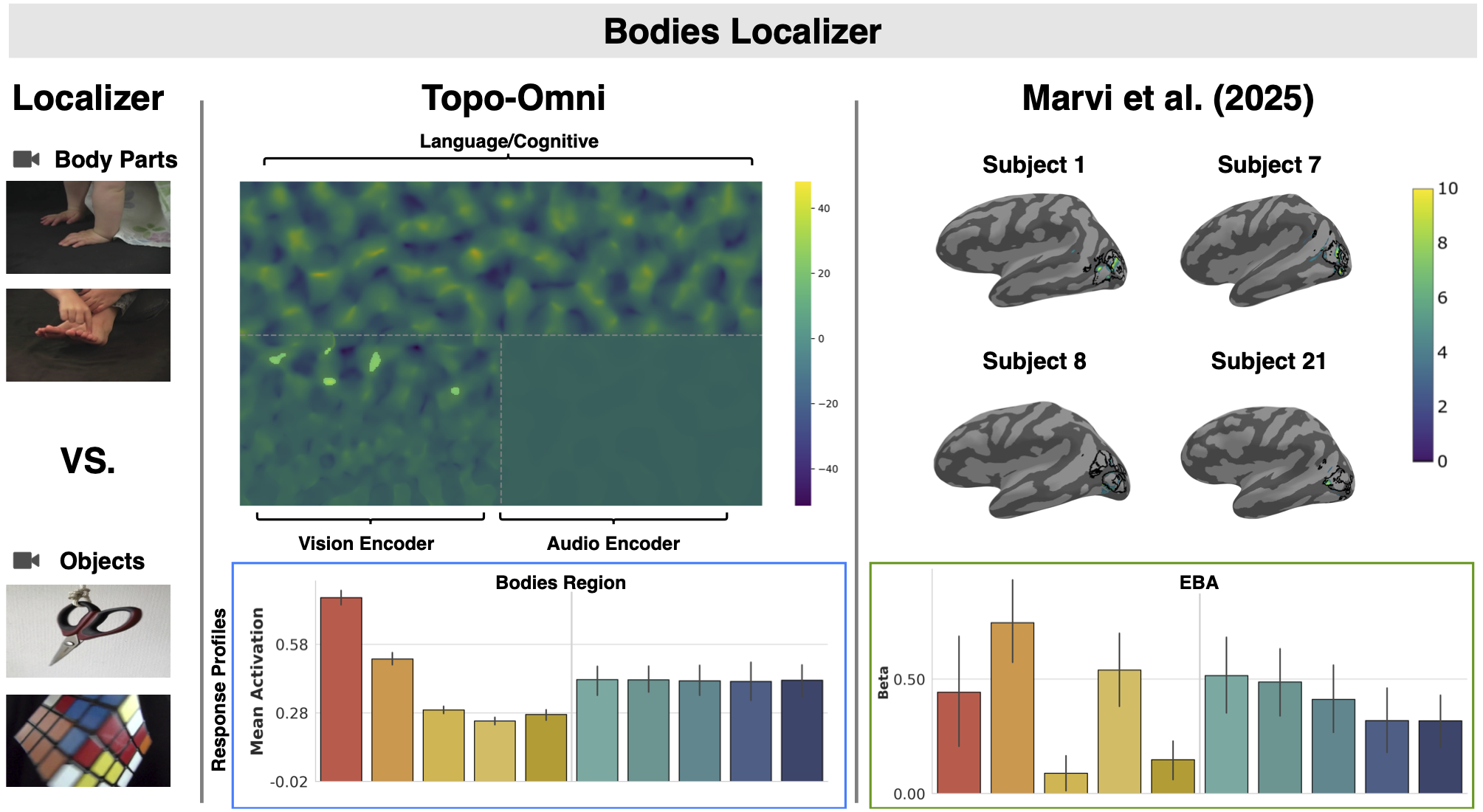}
\caption{
    \textbf{The bodies localizer isolates a body-selective region in the \ourmodel vision encoder that spatially parallels the extrastriate body area (EBA), but whose response profile does not significantly match human EBA.} Bodies localizer (Body Parts vs.\ Objects): in-silico (center) and human fMRI (right; $n=4$ subjects, from \citealp{Marvi2025}) activation maps, with yellow/green indicating contrast selectivity and anatomical-localizer clusters outlined. Response profiles (bottom) average across the top-1\% of model selective units (Bodies Region) and human EBA.
}
\label{fig:bodies-localizer}
\end{figure}

We additionally applied a body localizer (Body Parts vs.\ Objects; \citealp{Marvi2025}) to \ourmodel (Fig.~\ref{fig:bodies-localizer}). A spatially coherent body-selective cluster emerged in the vision encoder, responding preferentially to bodies---and, to a similar degree, faces---over the remaining categories (body-selectivity $d'=0.21$, paired $t(308)=26.4$, $p<0.001$, $n=309$ units). Unlike the other visual localizers, however, the region's full response profile was not significantly correlated with the human extrastriate body area (EBA) profile (Pearson $r=0.38$, $p=0.29$; Spearman $\rho=0.31$, $p=0.39$; permutation tests). Thus, while \ourmodel recovers a spatially organized, body-preferring region, its fine-grained tuning across categories diverges from human EBA. This dissociation, reliable category selectivity without a matching cross-category profile, likely reflects the modest body selectivity ($d'=0.21$). For the cortical sheet visualization of Figs.~\ref{fig:vision-clusters}, \ref{fig:audio-clusters}, \ref{fig:cognition-clusters} and \ref{fig:bodies-localizer} we are using a Gaussian kernel (FWHM $= 8.0$\,mm) to approximate the spatial sampling of fMRI.

\subsection{Response-profile correspondence and the effect of topography}
\label{sec:appendix:response-profiles}

\begin{figure}[H]
\centering
\includegraphics[width=1\textwidth]{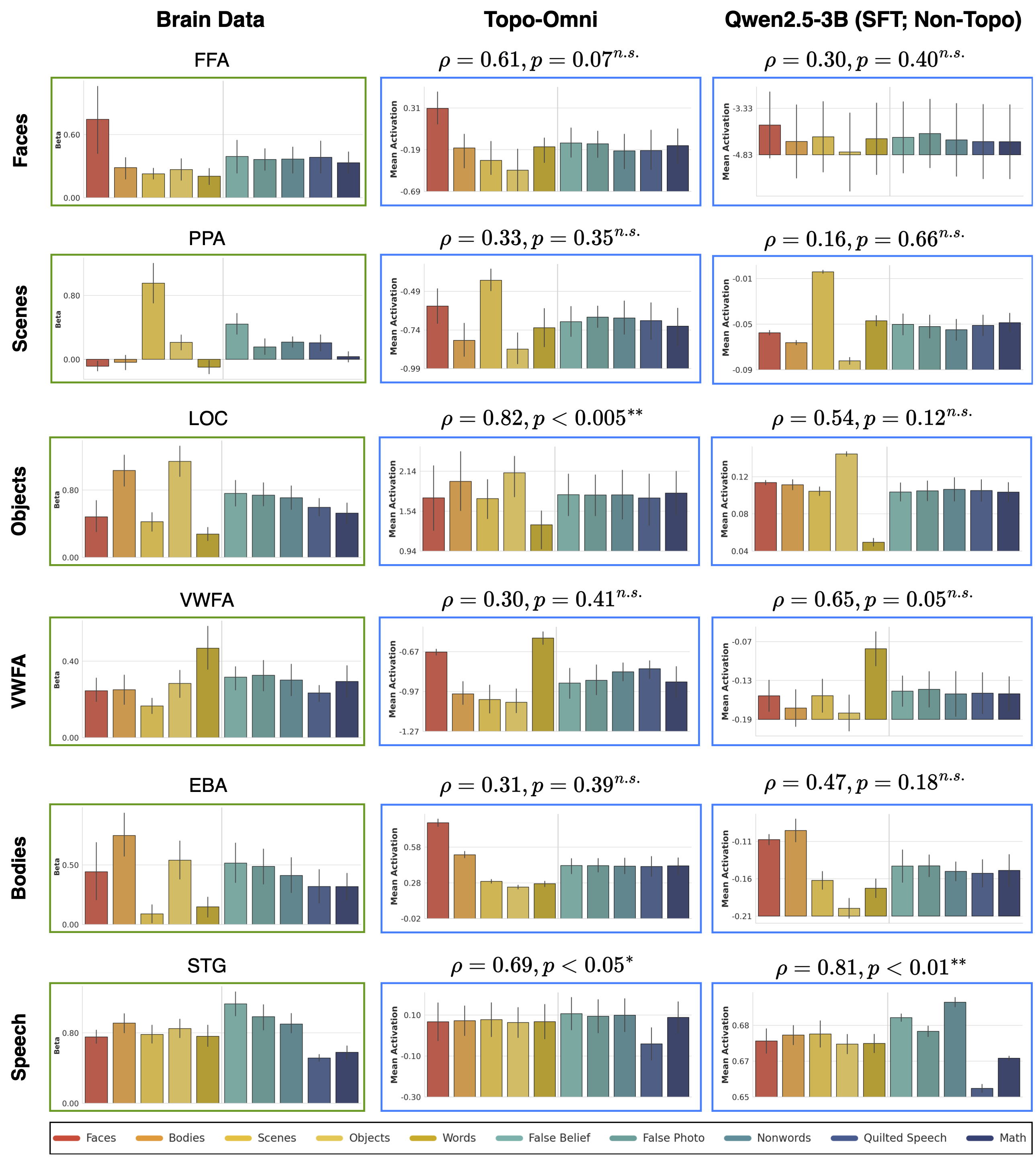}
\caption{
    \textbf{Response profiles of \ourmodel and a non-topographic baseline correlate with human ROI profiles to a comparable degree.}
    Each row shows the mean response profile across the ten stimulus conditions for one human ROI (left; brain data), the matched \ourmodel region (center), and the matched region of a non-topographic Qwen2.5-3B SFT baseline (right). Spearman correlations between each model profile and the human profile appear above each panel ($^{*}p<0.05$, $^{**}p<0.01$, $^{n.s.}$ not significant; permutation tests). Error bars are across subjects for the brain data and across units for the models.
}
\label{fig:more-response-profiles}
\end{figure}

To test whether the topographic objective alters the functional tuning that underlies brain similarity, we compared \ourmodel against a non-topographic Qwen2.5-Omni-3B SFT baseline trained without $\mathcal{L}_{\text{spatial}}$ (Fig.~\ref{fig:more-response-profiles}). For each human ROI we identified the matched region in each model---a spatial cluster in \ourmodel and the corresponding set of selective units in the baseline, which lacks spatial organization---and correlated its response profile across the ten stimulus conditions with the human profile (Spearman; permutation tests).
Both models reproduced the broad shape of most ROI profiles, with correlations of  comparable magnitude (mean $\rho=0.51$ for \ourmodel, $0.49$ for the baseline).\ourmodel reached significance for LOC ($\rho=0.82$, $p<0.005$) and STG ($\rho=0.69$, $p<0.05$), and the baseline for STG ($\rho=0.81$, $p<0.01$); the remaining ROIs showed positive but non-significant correlations for both models, consistent with the limited power of a ten-condition profile correlation. Neither model systematically outperformed the other: \ourmodel showed higher correlations for FFA, PPA, and LOC, and the baseline for VWFA, EBA, and STG. Adding the spatial smoothness term thus neither improved nor degraded response-profile correspondence, indicating that $\mathcal{L}_{\text{spatial}}$ reorganizes these functional responses across the cortical sheet without distorting the underlying tuning. The contribution of the topographic objective is therefore the emergent spatial organization documented in the main text: category-selective maps, retinotopy, tonotopy, and anatomically targeted interventions, obtained at no cost to the functional brain-similarity of the representations.

\end{appendices}

\end{document}